\documentclass[twocolumn,showpacs,showkeys,preprintnumbers,amssymb,aps,superscriptaddress,prb]{revtex4}
\usepackage{graphicx}
\usepackage{dcolumn}
\usepackage{color}
\usepackage{bm}

\begin{document}


\title{Diversity of quantum ground states and quantum phase transitions of a spin-1/2 Heisenberg octahedral chain}
\author{Jozef Stre\v{c}ka}
\email{jozef.strecka@upjs.sk}
\affiliation{Institute of Physics, Faculty of Science, P. J. \v{S}af\'{a}rik University, Park Angelinum 9, 04001 Ko\v{s}ice, Slovakia}
\author{Johannes Richter}
\affiliation{Institut f\"ur Theoretische Physik, Otto-von-Guericke Universit\"at in Magdeburg, 39016 Magdeburg, Germany}
\author{Oleg Derzhko}
\affiliation{Institute for Condensed Matter Physics, NASU, Svientsitskii Street 1, 79011 L'viv, Ukraine}
\affiliation{Department for Theoretical Physics, Ivan Franko National University of L'viv, Drahomanov Street 12, 79005 L'viv, Ukraine}
\author{Taras Verkholyak}
\affiliation{Institute for Condensed Matter Physics, NASU, Svientsitskii Street 1, 79011 L'viv, Ukraine}
\author{Katar\'ina Kar\v{l}ov\'a}
\affiliation{Institute of Physics, Faculty of Science, P. J. \v{S}af\'{a}rik University, Park Angelinum 9, 04001 Ko\v{s}ice, Slovakia}

\date{\today}

\begin{abstract}
The spin-$\frac{1}{2}$ Heisenberg octahedral chain with regularly alternating monomeric and square-plaquette sites is investigated using various analytical and numerical methods: variational technique, localized-magnon approach, exact diagonalization (ED) and density-matrix renormalization group (DMRG) method. The model belongs to the class of flat-band systems and it has a rich ground-state phase diagram including phases with spontaneously broken translational symmetry. Moreover, it exhibits an anomalous low-temperature thermodynamics close to continuous or discontinuous field-driven quantum phase transitions between three quantum ferrimagnetic phases, tetramer-hexamer phase, monomer-tetramer phase, localized-magnon phase and two different spin-liquid phases. If the intra-plaquette coupling is at least twice as strong as the monomer-plaquette coupling, the variational method furnishes a rigorous proof of the monomer-tetramer ground state in a low-field region and the localized-magnon approach provides an exact evidence of a single magnon trapped at each square plaquette in a high-field region. In the rest of parameter space we have numerically studied the ground-state phase diagram and magnetization process using DMRG and ED methods. It is shown that the zero-temperature magnetization curve may involve up to four intermediate plateaus at zero, one-fifth, two-fifth and three-fifth of the saturation magnetization, while the specific heat exhibits a striking low-temperature peak in a vicinity of discontinuous quantum phase transitions. 
\end{abstract}

\maketitle

\section{Introduction}

Quantum phase transitions belong to the most notable manifestations of low-dimensional quantum spin systems, which can be achieved upon variation of some external force like for instance magnetic field, mechanical or chemical pressure (doping).\cite{sach11} The quantum Heisenberg model exhibits a great diversity of unconventional quantum orders such as topologically ordered Haldane-type phases \cite{hald83} or dimerized states with an outstanding valence-bond-crystal order.\cite{miya03,miya11} Among the most notable disordered quantum states without any local order parameter one could further mention resonating-valence-bond phases \cite{ande73} or other types (e.g. Tomonaga-Luttinger) of quantum spin liquids.\cite{lhui02,bale10,misg11} A great variety of exotic quantum ground states can be found first of all in frustrated Heisenberg spin models due to a mutual interplay between quantum effects with a geometric spin frustration.
\cite{diep04,LNP2004,lacr11} 

The dimerized states with a valence-bond-crystal order are historically the most famous ground states of the frustrated spin-$\frac{1}{2}$ Heisenberg models, which arise out from an effort of antiferromagnetically coupled spins to create a singlet dimer (valence bond). Although the full exact solution of the Heisenberg spin models is usually beyond the scope of the present knowledge, the variational principle provides an efficient tool for a rigorous determination of the dimerized ground states for a few paradigmatic examples such as the Majumdar-Ghosh model,\cite{maju69,maju70} 
the Shastry-Sutherland model,\cite{shas81} the frustrated ladder,\cite{gelf91,hone00,chan06,hon2016} etc. However, the usage of variational arguments is regrettably restricted mostly to the highly-frustrated parameter region and low magnetic fields, while the Heisenberg spin systems often display spectacular quantum ground states also outside of this parameter space. For instance, the spin-$\frac{1}{2}$ Heisenberg diamond chain \cite{taka96} and diamond-like decorated planar lattices \cite{mori16,hiro16,hiro17} exhibit at moderate values of the spin frustration a peculiar tetramer-dimer ground state with spontaneously broken symmetry before they finally enter the monomer-dimer ground state predicted by the variational method in the highly-frustrated region.    

On the other hand, in flat-band quantum spin systems the concept of localized magnons \cite{schu02} affords a powerful tool for a rigorous assignment of quantum ground states of the geometrically frustrated Heisenberg spin models at sufficiently high magnetic fields (see Refs. \onlinecite{zhit05,derz06,derz15} for recent reviews and Refs. \onlinecite{tasa98,tasi98,gbjo12,mukh15,weim16,leyk13,flac14} for other flat-band systems). This technique can be employed whenever destructive quantum interference traps magnon(s) within cells with even number of bonds and hence, the frustrated quantum Heisenberg model can be exactly mapped onto a classical lattice-gas model with a hard-core potential.\cite{zhit05,derz06,derz15} Using this approach, the  microscopic nature of the last intermediate plateau in a zero-temperature magnetization curve of the quantum spin-$\frac{1}{2}$ Heisenberg kagom\'e lattice has been for instance elucidated along with a precise nature of a relevant second-order phase transition emerging at low but nonzero temperatures.\cite{zhit05} The main advantage of the localized-magnon approach lies in that it also provides, besides an exact ground state, accurate description of low-temperature thermodynamics due to a proper counting of low-lying excited states.\cite{zhit05,derz06,derz15}  

In the present work we will explore ground states, quantum phase transitions and low-temperature thermodynamics of the quantum spin-$\frac{1}{2}$ Heisenberg octahedral chain, in which quantum spins placed at monomeric sites regularly alternate with the ones residing square plaquettes (see Fig. \ref{fig1}). The proposed model belongs to a valuable class of the frustrated quantum Heisenberg models, which satisfy a local conservation of the total spin on square plaquettes. From this point of view, the spin-$\frac{1}{2}$ Heisenberg octahedral chain is quite reminiscent of the spin-$\frac{1}{2}$ Heisenberg diamond chain, which has been thoroughly investigated in relation with a frustrated magnetism of several copper-based magnetic compounds such as Cu$_3$(CO$_3$)$_2$(OH)$_2$ \cite{jesc11,hone11} and A$_3$Cu$_3$AlO$_2$(SO$_4$)$_4$ (A=K, Rb and Cs).\cite{fuji15,mori17} Apart from a few exact results to be obtained within the variational and localized-magnon approaches, the spin-$\frac{1}{2}$ Heisenberg octahedral chain can be rigorously mapped onto effective mixed-spin Heisenberg chains by following the approach developed previously by Honecker, Mila and Troyer.\cite{hone00,chan06} The DMRG simulations of the effective mixed-spin Heisenberg chains thus afford for the spin-$\frac{1}{2}$ Heisenberg octahedral chain precise numerical results, which will be additionally corroborated through the ED calculations.

It is worthwhile to remark that the polynuclear complexes, which involve quantum spin clusters with a geometric shape of octahedron as a magnetic core, constitute a relatively widespread family of compounds within an immense reservoir of coordination complexes. For illustration, let us quote a few specific examples of the hexanuclear complexes with an octahedral architecture of the magnetic core like Cu$_6$,\cite{liu003,xian05,zhao15,gao016} V$_6$,\cite{dani05,dani09} Cr$_6$,\cite{tsug96} Co$_6$,\cite{hong91} Fe$_6$,\cite{brec00,mugu04} Mn$_6$,\cite{arom99,stam09} Mo$_6$,\cite{sait90,cind00} W$_6$,\cite{ziet86,craw01} Ru$_6$,\cite{eady80} Ir$_6$,\cite{perg88} and Ta$_6$.\cite{peri09} Although we are currently not aware of any experimental realization of the spin-$\frac{1}{2}$ Heisenberg octahedral chain, we hope that our exciting theoretical findings presented hereafter could be inspiring for a tailored design of a one-dimensional polymeric chain of corner-sharing octahedra built out from discrete hexanuclear entities such as Cu$_6$\cite{liu003,xian05,zhao15,gao016} or V$_6$.\cite{dani05,dani09} 

The organization of this paper is as follows. The quantum spin-$\frac{1}{2}$ Heisenberg octahedral chain is introduced in Sec. \ref{sec:model} along with basic steps of analytical and numerical methods used for its treatment. The most interesting results for the ground-state phase diagram, zero-temperature magnetization process and low-temperature thermodynamics are discussed in Sec. \ref{sec:result}. Finally, several concluding remarks and future outlooks are mentioned in Sec. \ref{sec:conc}. 

\section{Heisenberg octahedral chain}
\label{sec:model}

\begin{figure}
\begin{center}
\includegraphics[width=0.45\textwidth]{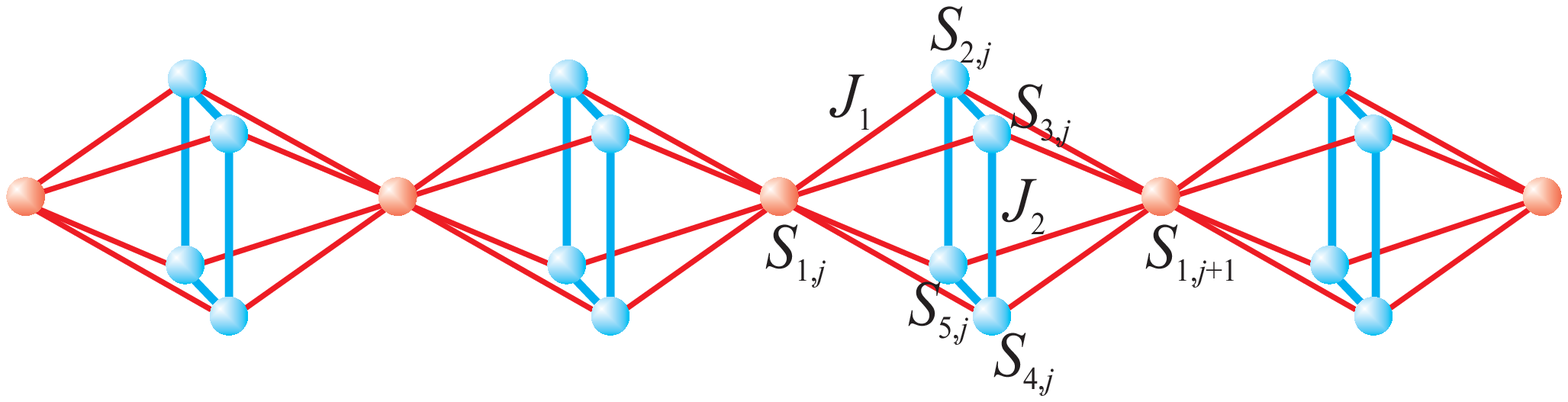}
\end{center}
\vspace{-0.6cm}
\caption{(Color online) A diagrammatic representation of the spin-$\frac{1}{2}$ Heisenberg octahedral chain. Thick (blue) lines represent the Heisenberg intra-plaquette coupling $J_2$, while thin (red) lines correspond to the monomer-plaquette coupling $J_1$.}
\label{fig1}
\end{figure}

Let us consider a one-dimensional chain of corner-sharing octahedra schematically depicted in Fig.~\ref{fig1}, which can be viewed as a generalization of the frustrated diamond chain \cite{taka96,jesc11,hone11} and the double-tetrahedra chain.\cite{mamb99,roja03,maks11} The Hamiltonian of the quantum spin-$\frac{1}{2}$ Heisenberg model defined upon the underlying octahedral chain is given by
\begin{eqnarray}
\label{ham}
\hat{\cal H} \!\!&=&\!\! 
\sum_{j=1}^{N} \Bigl[ J_1 (\boldsymbol{\hat{S}}_{1,j} + \boldsymbol{\hat{S}}_{1,j+1}) \!\cdot\! (\boldsymbol{\hat{S}}_{2,j} + \boldsymbol{\hat{S}}_{3,j} + \boldsymbol{\hat{S}}_{4,j} + \boldsymbol{\hat{S}}_{5,j}) \Bigr.  \nonumber \\
\!\!&+&\!\! J_2 (\boldsymbol{\hat{S}}_{2,j}\!\cdot\!\boldsymbol{\hat{S}}_{3,j} + \boldsymbol{\hat{S}}_{3,j}\!\cdot\!\boldsymbol{\hat{S}}_{4,j}
+ \boldsymbol{\hat{S}}_{4,j}\!\cdot\!\boldsymbol{\hat{S}}_{5,j} + \boldsymbol{\hat{S}}_{5,j}\!\cdot\!\boldsymbol{\hat{S}}_{2,j}) \nonumber \\
\Bigl. \!\!&-&\!\! h \sum_{i=1}^{5} \hat{S}_{i,j}^{z} \Bigr].
\end{eqnarray}
Above, $\boldsymbol{\hat{S}}_{i,j} \equiv (\hat{S}_{i,j}^x, \hat{S}_{i,j}^y, \hat{S}_{i,j}^z)$ denotes a standard spin-$\frac{1}{2}$ operator at a lattice site whose position is unambiguously determined by two subscripts, the former one specifies a position within the unit cell and the latter one the unit cell itself (see Fig.~\ref{fig2}). The coupling constant $J_1$ labels  the Heisenberg interaction between nearest-neighbor spins from monomeric and square-plaquette sites to be further referred to as the monomer-plaquette interaction, the coupling constant $J_2$ stands for the Heisenberg interaction between nearest-neighbor spins from the same square plaquette to be referred to as the intra-plaquette interaction and the Zeeman's term $h \geq 0$ accounts for a magnetostatic energy of magnetic moments in an external magnetic field. For simplicity, the periodic boundary condition $\boldsymbol{S}_{1,N+1} \equiv \boldsymbol{S}_{1,1}$ is imposed. The Hamiltonian (\ref{ham}) can be attacked by making use of several complementary analytical and numerical approaches, which will be dealt with in what follows. 

\subsection{Variational method}
\label{vm} 

\begin{figure}
\begin{center}
\includegraphics[width=0.1\textwidth]{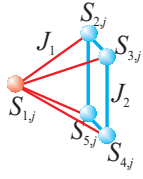}
\end{center}
\vspace{-0.6cm}
\caption{(Color online) The unit cell of the spin-$\frac{1}{2}$ Heisenberg octahedral chain, which is constituted by a five-spin cluster with the geometric shape of a square pyramid.}
\label{fig2}
\end{figure}

An exact ground state of the spin-$\frac{1}{2}$ Heisenberg octahedral chain can be rigorously found in the highly frustrated parameter region $J_2>2J_1$ and low enough magnetic fields $h<J_1+J_2$ by exploiting the variational principle.\cite{shas81,bose89,bose90,bose92} To this end, let us decompose the total Hamiltonian (\ref{ham}) of the spin-$\frac{1}{2}$ Heisenberg octahedral chain into a sum over the cell Hamiltonians 
\begin{eqnarray}
\hat{\cal H} = \sum_{j=1}^{N} \sum_{k=0}^{1} \hat{\cal H}_{j,k},
\label{hamsum}
\end{eqnarray}
whereas the cell Hamiltonian $\hat{\cal H}_{j,k}$ involves the interaction terms related to a five-spin cluster forming the unit cell with 
a geometric shape of a square pyramid (see Fig.~\ref{fig2})
\begin{eqnarray}
\hat{\cal H}_{j,k} \!\!&=&\!\! J_1 \boldsymbol{\hat{S}}_{1,j+k} \cdot (\boldsymbol{\hat{S}}_{2,j} + \boldsymbol{\hat{S}}_{3,j} 
                                 + \boldsymbol{\hat{S}}_{4,j} + \boldsymbol{\hat{S}}_{5,j}) \nonumber \\
                  \!\!&+&\!\!  \frac{J_2}{2} (\boldsymbol{\hat{S}}_{2,j}\!\cdot\!\boldsymbol{\hat{S}}_{3,j} + \boldsymbol{\hat{S}}_{3,j}\!\cdot\!\boldsymbol{\hat{S}}_{4,j}
+ \boldsymbol{\hat{S}}_{4,j}\!\cdot\!\boldsymbol{\hat{S}}_{5,j} + \boldsymbol{\hat{S}}_{5,j}\!\cdot\!\boldsymbol{\hat{S}}_{2,j}) \nonumber \\
 \!\!&-&\!\! \frac{h}{2} \sum_{i=1}^{5} \hat{S}_{i,j}^{z}.
\label{hamclu}
\end{eqnarray}
Note that the factor $\frac{1}{2}$ at two latter interaction terms avoids a double counting of the intra-plaquette coupling $J_2$ and the Zeeman's term $h$, which are symmetrically split into two consecutive cell Hamiltonians. The variational procedure enables one to obtain the lower bound for the ground-state energy $E_0$ of the spin-$\frac{1}{2}$ Heisenberg octahedral chain
\begin{eqnarray}
E_0 \!=\! \langle \Psi_0 | \hat{\cal H} | \Psi_0 \rangle \!=\! \langle \Psi_0 | \sum_{j=1}^{N} \sum_{k=0}^{1} \! \hat{\cal H}_{j,k} | \Psi_0 \rangle 
\!\geq\! \sum_{j=1}^{N} \sum_{k=0}^{1} \! \varepsilon_{j,k}^0, 
\label{var}
\end{eqnarray}
because the ground-state eigenvector $| \Psi_0 \rangle$ can be alternatively viewed as a variational function for the five-spin Heisenberg clusters (Fig.~\ref{fig2}). It follows from Eq. (\ref{var}) that the relevant ground-state energy $E_0$ must be necessarily greater or equal to the sum of the lowest-energy eigenenergies of the five-spin clusters $\varepsilon_{j,k}^0$. The energy spectrum of the five-spin Heisenberg cluster (Fig. \ref{fig2}) with the geometric arrangement of a square pyramid can be expressed in terms of five quantum spin numbers $S_{T,j,k}$, $S_{T,j,k}^{z}$, $S_{\square,j}$, $S_{24,j}$ and $S_{35,j}$
\begin{eqnarray}
\varepsilon_{j,k} \!\!&=&\!\! \frac{J_1}{2} S_{T,j,k}(S_{T,j,k}+1) + \!\left(\!\frac{J_2}{4}- \frac{J_1}{2}\!\right)\! S_{\square,j} (S_{\square,j} + 1) \nonumber \\
    \!\!&-&\!\! \frac{J_2}{4} [S_{24,j} (S_{24,j} + 1) + S_{35,j} (S_{35,j} + 1)] \nonumber \\
    \!\!&-&\!\! \frac{3}{8} J_1 - h S_{T,j,k}^{z},
\label{spec}
\end{eqnarray}
which determine the total spin of the square pyramid $S_{T,j,k}$ and its $z$-component $S_{T,j,k}^{z}$, the total spin of the square plaquette $S_{\square,j}$ and the total spin of two spin pairs from opposite corners of a square plaquette $S_{24,j}$ and $S_{35,j}$, respectively. It can be easily checked from Eq. (\ref{spec}) that the lowest-energy eigenstate of the five-spin Heisenberg cluster in the parameter space $h<J_1+J_2$ and $J_2>2J_1$ is a doublet state, which can be characterized by the following quantum spin numbers $S_{T,j,k} = |S_{T,j,k}^{z}| = \frac{1}{2}$, $S_{\square,j} = 0$, $S_{24,j}= 1$ and $S_{35,j} = 1$. Apparently, the four spins from each square plaquette are in a singlet-tetramer state given by the eigenvector
\begin{eqnarray}
|0,1,1\rangle_j \!\!\!&=&\!\!\! |S_{\square,j}=0, S_{24,j} = 1, S_{35,j}=1 \rangle   \nonumber \\
\!\!\!&=&\!\!\! \!\frac{1}{\sqrt{3}}(|\!\!\uparrow_{2,j}\downarrow_{3,j}\uparrow_{4,j}\downarrow_{5,j}\rangle + |\!\!\downarrow_{2,j}\uparrow_{3,j}\downarrow_{4,j}\uparrow_{5,j}\rangle)  \nonumber \\
\!\!\!&-&\!\!\! \frac{1}{\sqrt{12}} (|\!\!\uparrow_{2,j}\uparrow_{3,j}\downarrow_{4,j}\downarrow_{5,j}\rangle + |\!\!\uparrow_{2,j}\downarrow_{3,j}\downarrow_{4,j}\uparrow_{5,j}\rangle \nonumber \\
\!\!\!&+&\!\!\! |\!\!\downarrow_{2,j}\uparrow_{3,j}\uparrow_{4,j}\downarrow_{5,j}\rangle + 
|\!\!\downarrow_{2,j}\downarrow_{3,j}\uparrow_{4,j}\uparrow_{5,j}\rangle)\!
\label{ST}
\end{eqnarray}
and the spins from the monomeric sites are consequently decoupled from the other spins. This lowest-energy eigenstate can be readily extended to the whole spin-$\frac{1}{2}$ Heisenberg octahedral chain, which results in the monomer-tetramer (MT) ground state 
\begin{eqnarray}
|{\rm MT} \rangle \!\!=\!\! \prod_{j=1}^N \! |\!\!\uparrow_{1,j}\rangle \!\otimes\! 
\Bigl[\!\frac{1}{\sqrt{3}}(|\!\!\uparrow_{2,j}\downarrow_{3,j}\uparrow_{4,j}\downarrow_{5,j}\rangle \!\!\!&+&\!\!\! |\!\!\downarrow_{2,j}\uparrow_{3,j}\downarrow_{4,j}\uparrow_{5,j}\rangle)  \nonumber \\
- \frac{1}{\sqrt{12}} (|\!\!\uparrow_{2,j}\uparrow_{3,j}\downarrow_{4,j}\downarrow_{5,j}\rangle \!\!\!&+&\!\!\! |\!\!\uparrow_{2,j}\downarrow_{3,j}\downarrow_{4,j}\uparrow_{5,j}\rangle \nonumber \\
+ |\!\!\downarrow_{2,j}\uparrow_{3,j}\uparrow_{4,j}\downarrow_{5,j}\rangle \!\!\!&+&\!\!\! |\!\!\downarrow_{2,j}\downarrow_{3,j}\uparrow_{4,j}\uparrow_{5,j}\rangle) \Bigr]\!. \nonumber \\
\label{MT}
\end{eqnarray}
To conclude this part, the spin-$\frac{1}{2}$ Heisenberg octahedral chain exhibits the exact MT ground state in the low-field part $h<J_1+J_2$ of the highly frustrated parameter region $J_2>2J_1$, where four spins from each square plaquette form a singlet-tetramer state and the spins from the 
monomeric sites are fully aligned into the magnetic field or they are completely free to flip in a zero field. With regard to the perfect alignment of all monomeric spins, the MT ground state should manifest itself as the intermediate one-fifth plateau present in a zero-temperature magnetization curve within the field range $h \in (0, J_1 + J_2)$.

\subsection{Localized-magnon approach: ground state}
\label{lmgs}

\begin{figure}
\begin{center}
\includegraphics[width=0.5\textwidth]{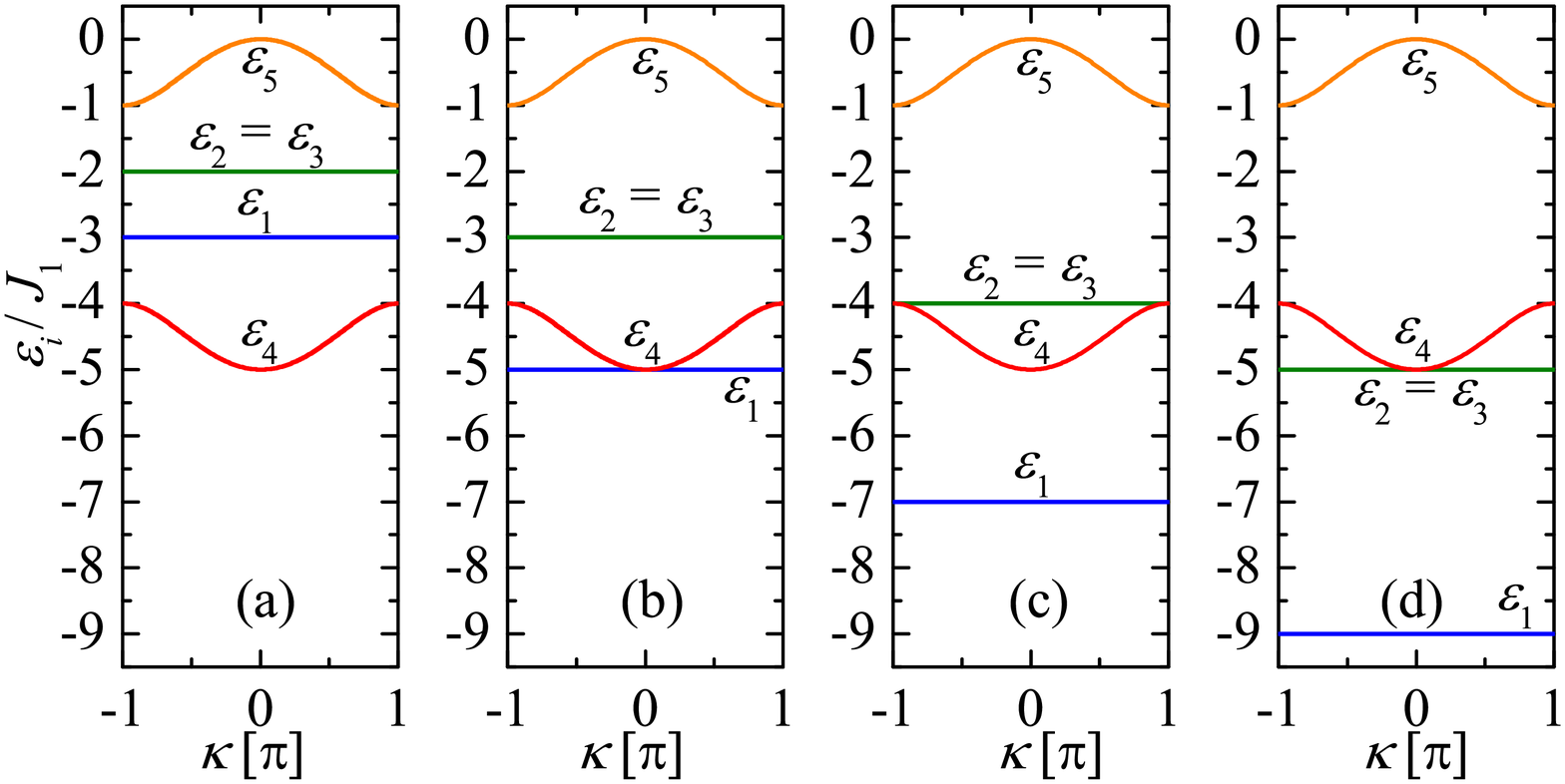}
\end{center}
\vspace{-0.6cm}
\caption{(Color online) The one-magnon energy bands (\ref{oms}) of the spin-$\frac{1}{2}$ Heisenberg octahedral chain  for four different values of the interaction ratio: (a) $J_2/J_1 = 1$; (b) $J_2/J_1 = 2$; (c) $J_2/J_1 = 3$; (d) $J_2/J_1 = 4$.}
\label{fig3}
\end{figure}

It is quite clear that the lowest-energy eigenstate of the spin-$\frac{1}{2}$ Heisenberg octahedral chain at sufficiently high magnetic fields exceeding the 
saturation value is the fully polarized ferromagnetic (FM) state 
\begin{eqnarray}
|{\rm FM} \rangle = \prod_{j=1}^N \! |\!\!\uparrow_{1,j}\uparrow_{2,j}\uparrow_{3,j}\uparrow_{4,j}\uparrow_{5,j}\rangle 
\label{FM}
\end{eqnarray}
with the following energy eigenvalue $E_{\rm FM} = E_{\rm FM}^0 - \frac{5N}{2} h$, $E_{\rm FM}^0 = N (2J_1 + J_2)$ is the respective zero-field energy. It will be demonstrated hereafter that the concept of independent localized magnons \cite{derz06,derz15} can be employed in the highly frustrated region $J_2>2J_1$ for a rigorous assignment of the 
saturation field and the exact ground state emerging below the saturation field. The one-magnon eigenstates can be constructed within the orthonormal basis set $|i, j \rangle = \hat{S}_{i,j}^{-} |{\rm FM} \rangle$ ($i=1-5$, $j=1-N$) belonging to the sector $S_T^z = \frac{5N}{2} - 1$ with a single spin deviation from the fully polarized FM state. Applying the zero-field part of the Hamiltonian (\ref{ham}) within the given basis leads to the following set of equations
\begin{eqnarray}
\hat{\cal H} |1, j\rangle \!\!\!&=&\!\!\! (E_{\rm FM}^0 - 4 J_1) |1, j\rangle  + \frac{J_1}{2} \sum_{i=2}^5 (|i, j-1\rangle + |i, j\rangle), \nonumber \\
\hat{\cal H} |2, j\rangle \!\!\!&=&\!\!\! (E_{\rm FM}^0 - J_1 - J_2) |2, j\rangle  + \frac{J_1}{2} (|1, j\rangle + |1, j+1\rangle) \nonumber \\
\!\!\!&+&\!\!\! \frac{J_2}{2} (|3, j\rangle + |5, j\rangle), \nonumber \\
\hat{\cal H} |3, j\rangle \!\!\!&=&\!\!\! (E_{\rm FM}^0 - J_1 - J_2) |3, j\rangle  + \frac{J_1}{2} (|1, j\rangle + |1, j+1\rangle) \nonumber \\
\!\!\!&+&\!\!\! \frac{J_2}{2} (|2, j\rangle + |4, j\rangle), \nonumber \\
\hat{\cal H} |4, j\rangle \!\!\!&=&\!\!\! (E_{\rm FM}^0 - J_1 - J_2) |4, j\rangle  + \frac{J_1}{2} (|1, j\rangle + |1, j+1\rangle) \nonumber \\
\!\!\!&+&\!\!\! \frac{J_2}{2} (|3, j\rangle + |5, j\rangle), \nonumber \\
\hat{\cal H} |5, j\rangle \!\!\!&=&\!\!\! (E_{\rm FM}^0 - J_1 - J_2) |5, j\rangle  + \frac{J_1}{2} (|1, j\rangle + |1, j+1\rangle) \nonumber \\
\!\!\!&+&\!\!\! \frac{J_2}{2} (|2, j\rangle + |4, j\rangle), 
\label{lm}
\end{eqnarray}
which can be used for solving the eigenvalue problem $\hat{\cal H} |\Psi_k\rangle = E_k^0 |\Psi_k\rangle$ in a zero field within the one-magnon sector by assuming $|\Psi_k\rangle = \sum_{i=1}^{5} \sum_{j=1}^{N} c_{i,\kappa} {\rm e}^{{\rm i} \kappa j}|i, j \rangle$. The solution of the eigenvalue problem follows from the characteristic equation 
\begin{widetext}
\begin{eqnarray}
\left|
\begin{array}{ccccc}
-4J_1-\varepsilon_k & \frac{J_1}{2} (1 + {\rm e}^{-{\rm i} \kappa}) & \frac{J_1}{2} (1 + {\rm e}^{-{\rm i} \kappa}) 
& \frac{J_1}{2} (1 + {\rm e}^{-{\rm i} \kappa}) & \frac{J_1}{2} (1 + {\rm e}^{-{\rm i} \kappa}) \\ 	
\frac{J_1}{2} (1 + {\rm e}^{{\rm i} \kappa}) & -J_1-J_2-\varepsilon_k & \frac{J_2}{2} & 0 & \frac{J_2}{2} \\
\frac{J_1}{2} (1 + {\rm e}^{{\rm i} \kappa}) & \frac{J_2}{2} & -J_1-J_2-\varepsilon_k & \frac{J_2}{2} & 0 \\
\frac{J_1}{2} (1 + {\rm e}^{{\rm i} \kappa}) & 0 & \frac{J_2}{2} & -J_1-J_2-\varepsilon_k & \frac{J_2}{2} \\
\frac{J_1}{2} (1 + {\rm e}^{{\rm i} \kappa}) & \frac{J_2}{2} & 0 & \frac{J_2}{2} & -J_1-J_2-\varepsilon_k \\
\end{array}
\right| = 0,
\label{det}
\end{eqnarray}
\end{widetext}
where $\varepsilon_k = E_k^0 - E_{\rm FM}^0$ labels an energy difference between the one-magnon state and the fully polarized FM state in a zero magnetic field. The one-magnon energy spectrum of the spin-$\frac{1}{2}$ Heisenberg octahedral chain in a zero magnetic field is composed of five energy bands
\begin{eqnarray}
\varepsilon_{1}   \!\!\!&=&\!\!\! - J_1 - 2J_2,  \nonumber \\
\varepsilon_{2,3} \!\!\!&=&\!\!\! - J_1 -  J_2,  \nonumber \\
\varepsilon_{4,5} \!\!\!&=&\!\!\! - \frac{J_1}{2} \left(5 \pm \sqrt{17 + 8 \cos \kappa} \right),  
\label{oms}
\end{eqnarray}
which are for illustration depicted in Fig. \ref{fig3} for a few selected values of the interaction ratio $J_2/J_1$. It should be pointed out that three out of five one-magnon energy bands (\ref{oms}) are completely dispersionless (flat), which implies presence of the localized magnons within the flat bands.\cite{derz06,derz15} The flat band with the eigenenergy $\varepsilon_{1}$ supports a single magnon trapped within a square plaquette
\begin{eqnarray}
|lm\rangle_j = \frac{1}{2} \left(\hat{S}_{2,j}^{-} - \hat{S}_{3,j}^{-} + \hat{S}_{4,j}^{-} - \hat{S}_{5,j}^{-}\right) |{\rm FM}\rangle
\label{om1}
\end{eqnarray}
and this one-magnon state corresponding to the quantum spin numbers $S_{\square,j}=S^z_{\square,j}=S_{24,j}=S_{35,j}=1$ has the lowest energy in the highly frustrated regime $J_2 > 2 J_1$ (see Fig. \ref{fig3}). 

The many-magnon states of the spin-$\frac{1}{2}$ Heisenberg octahedral chain can be subsequently obtained by filling in square plaquettes with the localized magnons of the type (\ref{om1}). The eigenstates involving $N_1$ independent one-magnon states (\ref{om1}) trapped on square plaquettes have the energy $E_{N_1} = E_{\rm FM} - N_1 (|\varepsilon_{1}| - h)$, which implies the following value of the saturation field $h_s = |\varepsilon_{1}| = J_1 + 2 J_2$ in the highly frustrated region $J_2 > 2 J_1$. In addition, it can be easily verified that the lowest-energy state in the parameter space $J_2 > 2 J_1$ and $h<h_s$ is the many-magnon state with the highest possible number ($N$) of the independent localized magnons (\ref{om1}) on the square plaquettes 
\begin{eqnarray}
|{\rm LM}\rangle = \prod_{j=1}^N \! |\!\!\uparrow_{1,j}\rangle \!\otimes\! \frac{1}{2}
(|\!\!\downarrow_{2,j}\uparrow_{3,j}\uparrow_{4,j}\uparrow_{5,j}\rangle 
\!\!\!&-&\!\!\!|\!\!\uparrow_{2,j}\downarrow_{3,j}\uparrow_{4,j}\uparrow_{5,j}\rangle \nonumber \\
+|\!\!\uparrow_{2,j}\uparrow_{3,j}\downarrow_{4,j}\uparrow_{5,j}\rangle
\!\!\!&-&\!\!\!|\!\!\uparrow_{2,j}\uparrow_{3,j}\uparrow_{4,j}\downarrow_{5,j}\rangle). \nonumber \\  
\label{LM}
\end{eqnarray}
The localized-magnon ground state (\ref{LM}) should manifest itself in a zero-temperature magnetization curve as the intermediate three-fifth plateau restricted to the field range $h \in (J_1 + J_2, J_1 + 2J_2)$.

\subsection{Localized-magnon approach: thermodynamics} 
\label{lmt}

In the highly frustrated parameter space $J_2 > 2 J_1$ the concept of localized magnons \cite{derz06,derz15} can also be adapted for a rigorous description of low-temperature thermodynamics of the spin-$\frac{1}{2}$ Heisenberg octahedral chain. Under this circumstance, the many-magnon states constructed from the lowest-energy one-magnon state (\ref{om1}) are the 
most relevant low-lying states in the high-field region $h > J_1 + J_2$, while the many-magnon states including the localized two-magnon state (\ref{ST}) are the 
most important low-lying states in the low-field region $h < J_1 + J_2$. The low-temperature thermodynamics of the spin-$\frac{1}{2}$ Heisenberg octahedral chain can be accordingly reformulated as a two-component lattice-gas model, since each square plaquette can host at 
most one localized one-magnon state (\ref{om1}) represented by the first kind of particles with the chemical potential $\mu_1 = J_1 + 2J_2 - h$ or one localized two-magnon state (\ref{ST}) represented by the second kind of particles with the chemical potential $\mu_2 = 2J_1 + 3J_2 - 2h$. The chemical potentials $\mu_1$ and $\mu_2$ of two species of the particles are determined by an energy difference between the fully polarized ferromagnetic state (vacuum) and the respective localized magnon state (either one-magnon or two-magnon state). By introducing the occupation numbers $n_{1,j} = 0,1$ and $n_{2,j} = 0,1$ for the two species of the particles the overall energy of the many-magnon states is given by the classical lattice-gas Hamiltonian
\begin{eqnarray}
{\cal H} = E_{\rm FM} - \mu_1 \sum_{j=1}^{N} n_{1,j} - \mu_2 \sum_{j=1}^{N} n_{2,j}. \nonumber
\label{elm}
\end{eqnarray} 
The partition function of the spin-$\frac{1}{2}$ Heisenberg octahedral chain accounting for all available many-magnon states constituted from the lowest-energy one-magnon (\ref{om1}) and two-magnon (\ref{ST}) states then follows from the formula
\begin{eqnarray}
{\cal Z} \!\!&=&\!\! {\rm e}^{-\beta E_{\rm FM}} \prod_{j=1}^N  \sum_{n_{1,j}} \sum_{n_{2,j}} (1-n_{1,j}n_{2,j}) 
{\rm e}^{\beta (\mu_1 n_{1,j} + \mu_2 n_{2,j})} \nonumber \\
\!\!&=&\!\! {\rm e}^{-\beta E_{\rm FM}} \Bigl(1 + {\rm e}^{\beta \mu_1} + {\rm e}^{\beta \mu_2}\Bigr)^N\!\!\!\!,
\label{lmpf}
\end{eqnarray}
where $\beta = 1/(k_{\rm B} T)$, $k_{\rm B}$ is Boltzmann's constant, $T$ is the absolute temperature and the prefactor $(1-n_{1,j}n_{2,j})$ establishes a hard-core constraint for both kinds of the particles as each square plaquette can host at 
most one one-magnon state (\ref{om1}) or one two-magnon state (\ref{ST}) or should be kept empty provided that a square plaquette is fully polarized. The Helmholtz free energy per spin can be calculated from the relation 
\begin{eqnarray}
f = \!\!&-&\!\! k_{\rm B} T \lim_{N \to \infty} \frac{1}{5N} \ln {\cal Z} = \frac{1}{5} (2J_1 + J_2) - \frac{h}{2} \nonumber \\
\!\!&-&\!\! \frac{1}{5} k_{\rm B} T \ln \left(1 + {\rm e}^{\beta \mu_1} + {\rm e}^{\beta \mu_2} \right).
\label{lmgfe}
\end{eqnarray}
The Helmholtz free energy (\ref{lmgfe}) can be utilized for a calculation of the magnetization per spin
\begin{eqnarray}
m = \frac{1}{2} - \frac{1}{5} \frac{{\rm e}^{\beta \mu_1} + 2 {\rm e}^{\beta \mu_2}}{1 + {\rm e}^{\beta \mu_1} + {\rm e}^{\beta \mu_2}}
\label{lmmag}
\end{eqnarray}
and the specific heat per spin
\begin{eqnarray}
c = \frac{\mu_1^2 {\rm e}^{\beta \mu_1} + \mu_2^2 {\rm e}^{\beta \mu_2} 
          + (\mu_1 - \mu_2)^2 {\rm e}^{\beta (\mu_1 + \mu_2)}}
         {5 k_{\rm B} T^2 \left(1 + {\rm e}^{\beta \mu_1} + {\rm e}^{\beta \mu_2} \right)^2}.
\label{lmsh}
\end{eqnarray}
It is worthwhile to remark that the derived expressions for the Helmholtz free energy (\ref{lmgfe}), magnetization (\ref{lmmag}) and specific heat (\ref{lmsh}) provide valuable description of low-temperature thermodynamics just in the highly frustrated parameter space $J_2 > 2 J_1$. 

\subsection{Local conservation law and DMRG treatment of the effective mixed-spin chains} 

One of the most essential features of the spin-$\frac{1}{2}$ Heisenberg octahedral chain is being a local conservation of the total spin on square plaquettes, which directly follows from a validity of the commutation relation $[\hat{\cal H}, \boldsymbol{\hat{S}}_{\square, j}^{2}] = 0$ 
between the Hamiltonian (\ref{ham}) and the square of spin operator $\boldsymbol{\hat{S}}_{\square, j} = \boldsymbol{\hat{S}}_{2,j} + \boldsymbol{\hat{S}}_{3,j} + \boldsymbol{\hat{S}}_{4,j} + \boldsymbol{\hat{S}}_{5,j}$. It is therefore convenient to rewrite the zero-field part of the Hamiltonian (\ref{ham}) in terms of the total spin operator $\boldsymbol{\hat{S}}_{\square, j}$ of the square plaquette and two auxiliary spin operators $\boldsymbol{\hat{S}}_{24, j} = \boldsymbol{\hat{S}}_{2,j} + \boldsymbol{\hat{S}}_{4,j}$ and $\boldsymbol{\hat{S}}_{35, j} = \boldsymbol{\hat{S}}_{3, j} + \boldsymbol{\hat{S}}_{5, j}$ related to the spin pairs from opposite corners of square plaquettes (see Figs. \ref{fig1} and \ref{fig2})
\begin{eqnarray}
\hat{\cal H} \!\!&=&\!\! J_1 \sum_{j=1}^{N} (\boldsymbol{\hat{S}}_{1,j} + \boldsymbol{\hat{S}}_{1,j+1}) \!\cdot\! \boldsymbol{\hat{S}}_{\square, j} \nonumber \\  
\!\!&+&\!\! \frac{J_2}{2} \sum_{j=1}^{N} (\boldsymbol{\hat{S}}_{\square, j}^2 - \boldsymbol{\hat{S}}_{24, j}^2  - \boldsymbol{\hat{S}}_{35, j}^2). 
\label{hamlcl}
\end{eqnarray}
The effective Hamiltonian (\ref{hamlcl}) evidently corresponds to the ferrimagnetic mixed spin-$(\frac{1}{2},S_{\square,j})$ Heisenberg chains with some shift of energy eigenvalues due to quantum spin numbers $S_{\square,j}$, $S_{24,j}$ and $S_{35,j}$, whereas the quantum number determining the total spin on a square plaquette may achieve three different values $S_{\square,j} = 0$, $1$ and $2$. Hence, it follows that the ground state of the spin-$\frac{1}{2}$ Heisenberg octahedral chain can be found from the lowest-energy eigenstates of the effective Hamiltonian (\ref{hamlcl}) by assuming all possible combinations of the involved quantum spin numbers. On assumption that the translational period of a ground state is not broken one arrives just at three effective Hamiltonians corresponding to the fragmentized (paramagnetic) mixed spin-$\frac{1}{2}$ and spin-$0$ system
\begin{eqnarray}
\hat{\cal H}_{\frac{1}{2}-0} = -2 N J_2, 
\label{ham120}
\end{eqnarray}
the ferrimagnetic mixed spin-$(\frac{1}{2},1)$ Heisenberg chain
\begin{eqnarray}
\hat{\cal H}_{\frac{1}{2}-1} = J_1 \sum_{j=1}^{N} (\boldsymbol{\hat{S}}_{1,j} + \boldsymbol{\hat{S}}_{1,j+1}) 
                               \!\cdot\! \boldsymbol{\hat{S}}_{\square, j}  - N J_2, \,\,\,\, (S_{\square, j} = 1) \nonumber \\
\label{ham121}
\end{eqnarray}
and the ferrimagnetic mixed spin-$(\frac{1}{2},2)$ Heisenberg chain
\begin{eqnarray}
\hat{\cal H}_{\frac{1}{2}-2} = J_1 \sum_{j=1}^{N} (\boldsymbol{\hat{S}}_{1,j} + \boldsymbol{\hat{S}}_{1,j+1}) 
                               \!\cdot\! \boldsymbol{\hat{S}}_{\square, j}  + N J_2. \,\,\,\, (S_{\square, j} = 2) \nonumber \\ 
\label{ham122}
\end{eqnarray}
The lowest-energy eigenvalues of the effective Hamiltonians (\ref{ham120})-(\ref{ham122}) of the spin-$\frac{1}{2}$ Heisenberg octahedral chain then readily follow from
\begin{eqnarray}
E_{\frac{1}{2}-0} (2N, S_T^z) \!\!&=&\!\! -2 N J_2, \label{ee120} \\
E_{\frac{1}{2}-1} (2N, S_T^z) \!\!&=&\!\! 2 N J_1 \varepsilon_{\frac{1}{2}-1} (2N, S_T^z) - N J_2, \label{ee121} \\
E_{\frac{1}{2}-2} (2N, S_T^z) \!\!&=&\!\! 2 N J_1 \varepsilon_{\frac{1}{2}-2} (2N, S_T^z) + N J_2. \label{ee122}
\end{eqnarray}
Here, $\varepsilon_{\frac{1}{2}-S_{\square}} (2N, S_T^z)$ denotes the lowest-energy eigenvalue per spin of the mixed spin-$(\frac{1}{2},S_{\square})$ Heisenberg chain with the unit coupling constant and the total number of $2N$ spins in each sector with the $z$-component of the total spin $S_T^z$. The lowest-energy eigenvalue $E_{\frac{1}{2}-0} (2N, S_T^z) = -2 N J_2$ apparently corresponds to the monomer-tetramer ground state (\ref{MT}) with the paramagnetic character of the monomeric spins and all square plaquettes in the singlet-tetramer state (\ref{ST}), which is the true ground state for $J_2>2J_1$ and $h<J_1+J_2$ as exemplified by the variational method presented in Sect. \ref{vm}. On the contrary, the effective mixed spin-$(\frac{1}{2},S)$ Heisenberg chains with the uniform nearest-neighbor antiferromagnetic coupling exhibit at low enough magnetic fields the Lieb-Mattis ferrimagnetic ground state manifested in a zero-temperature magnetization curve as an intermediate plateau at $(2S-1)/(2S+1)$ of the saturation magnetization, which breaks down at a field-driven quantum phase transition towards the Tomonaga-Luttinger spin-liquid phase extending up to the saturation field.\cite{ivan98,yama99,saka99,yama00,saka02,teno11,stre17a,stre17b} Hence, one may expect in a magnetization process of the spin-$\frac{1}{2}$ Heisenberg octahedral chain emergence of the intermediate plateaus due to the Lieb-Mattis ferrimagnetic ground state as well as the gapless region inherent to the Tomonaga-Luttinger spin-liquid state.

However, one cannot exclude neither the possibility that the period of a ground state is spontaneously broken and the quantum spin number $S_{\square,j}$ determining the total spin of square plaquette varies along the effective mixed spin-$(\frac{1}{2},S_{\square,j})$ Heisenberg chain. We have therefore took into account the possible doubling of unit cell by considering another effective Hamiltonian of the ferrimagnetic mixed spin-$(\frac{1}{2},2,\frac{1}{2},1)$ Heisenberg chain 
\begin{eqnarray}
\hat{\cal H}_{\frac{1}{2}-2-\frac{1}{2}-1} = 
J_1 \sum_{j=1}^{N} (\boldsymbol{\hat{S}}_{1,j} + \boldsymbol{\hat{S}}_{1,j+1}) \!\cdot\! \boldsymbol{\hat{S}}_{\square, j}, 
\label{ham122121}
\end{eqnarray}
which assumes a regular alternation of the total spin $S_{\square, 2j-1} = 2$ and $S_{\square, 2j} = 1$ on odd and even square plaquettes. Another possible lowest-energy eigenvalue of the spin-$\frac{1}{2}$ Heisenberg octahedral chain may thus follow from the formula
\begin{eqnarray}
E_{\frac{1}{2}-2-\frac{1}{2}-1} (2N, S_T^z) \!\!&=&\!\! 2 N J_1 \varepsilon_{\frac{1}{2}-2-\frac{1}{2}-1} (2N, S_T^z), 
\label{eeeffd}
\end{eqnarray}
where $\varepsilon_{\frac{1}{2}-2-\frac{1}{2}-1} (2N, S_T^z)$ denotes the lowest-energy eigenvalue per spin of the mixed spin-$(\frac{1}{2},2,\frac{1}{2},1)$ Heisenberg chain with the unit coupling constant and the total number of $2N$ spins in each sector with the $z$-component of the total spin $S_T^z$. 

The lowest-energy eigenvalues $\varepsilon_{\frac{1}{2}-2} (2N, S_T^z)$, $\varepsilon_{\frac{1}{2}-1} (2N, S_T^z)$ and $\varepsilon_{\frac{1}{2}-2-\frac{1}{2}-1} (2N, S_T^z)$ of all aforedescribed effective mixed-spin Heisenberg chains with the total number of 120 spins ($N=60$) were calculated for all available sectors with the $z$-component of the total spin $S_T^z$ by means the numerical DMRG method when adapting the subroutine from Algorithms and Libraries for Physics Simulations (ALPS) project.\cite{baue11} It should be pointed out that the obtained numerical DMRG data correspond to the spin-$\frac{1}{2}$ Heisenberg octahedral chain with $N=60$ unit cells, i.e. $L=300$ spins.  

\begin{figure}
\begin{center}
\includegraphics[width=0.45\textwidth]{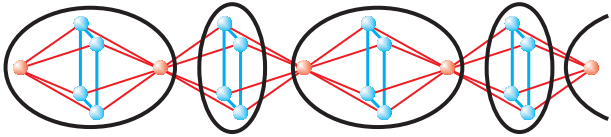}
\end{center}
\vspace{-0.6cm}
\caption{(Color online) A schematic representation of the singlet tetramer-hexamer state (\ref{SHT}). Thick (black) ovals represent singlet states of tetramers and hexamers given by Eqs. (\ref{ST}) and (\ref{SH}), respectively.}
\label{fig4}
\end{figure}

Last but not least, the regular alternation of the total spin of square plaquettes $S_{\square, 2j-1} = 1$ and $S_{\square, 2j} = 0$ leads to the  effective Hamiltonian of the ferrimagnetic mixed spin-$(\frac{1}{2},1,\frac{1}{2},0)$ Heisenberg chain, whose lowest-energy eigenstate can be found on analytical grounds because of a fragmentation at even square plaquettes in the singlet-tetramer state (\ref{ST}) with $S_{\square, 2j} = 0$. Owing to this fact, the effective mixed spin-$(\frac{1}{2},1,\frac{1}{2},0)$ Heisenberg chain decomposes into a set of the mixed spin-($\frac{1}{2},1,\frac{1}{2}$) Heisenberg trimers separated from each other by the non-magnetic spin-$0$ 
monomers. It can be easily verified that the lowest-energy eigenstate of the fragmentized mixed spin-$(\frac{1}{2},1,\frac{1}{2},0)$ Heisenberg chain is the singlet tetramer-hexamer state
\begin{eqnarray}
|{\rm TH} \rangle = \prod_{j=1}^{N/2} |0, 1, 1, 1, 1 \rangle_{2j-1} \otimes |0, 1, 1 \rangle_{2j}.
\label{SHT}
\end{eqnarray}
In above, the former state vector refers to the spin-$\frac{1}{2}$ Heisenberg octahedron in a singlet hexamer state
\begin{eqnarray}
|0, 1, 1, 1, 1 \rangle_{j} \!=\! |S_{T,j} = 0, S_{\square,j} \!\!\!&=&\!\!\! S_{24,j} \!=\! S_{35,j} \!=\! S_{16,j} \!=\! 1 \rangle \!=\!  \nonumber \\ \frac{1}{\sqrt{12}}  \Bigl   
(|\!\!\uparrow_{1,j}\uparrow_{2,j}\downarrow_{3,j}\uparrow_{4,j}\downarrow_{5,j}\downarrow_{6,j}\rangle 
\!\!\!&+&\!\!\!|\!\!\downarrow_{1,j}\uparrow_{2,j}\downarrow_{3,j}\uparrow_{4,j}\downarrow_{5,j}\uparrow_{6,j}\rangle \nonumber \\
+|\!\!\uparrow_{1,j}\downarrow_{2,j}\uparrow_{3,j}\downarrow_{4,j}\downarrow_{5,j}\uparrow_{6,j}\rangle 
\!\!\!&+&\!\!\!|\!\!\uparrow_{1,j}\downarrow_{2,j}\downarrow_{3,j}\downarrow_{4,j}\uparrow_{5,j}\uparrow_{6,j}\rangle \nonumber \\
+|\!\!\downarrow_{1,j}\uparrow_{2,j}\uparrow_{3,j}\downarrow_{4,j}\uparrow_{5,j}\downarrow_{6,j}\rangle  
\!\!\!&+&\!\!\!|\!\!\downarrow_{1,j}\downarrow_{2,j}\uparrow_{3,j}\uparrow_{4,j}\uparrow_{5,j}\downarrow_{6,j}\rangle \nonumber \\
-|\!\!\uparrow_{1,j}\downarrow_{2,j}\uparrow_{3,j}\downarrow_{4,j}\uparrow_{5,j}\downarrow_{6,j}\rangle 
\!\!\!&-&\!\!\!|\!\!\downarrow_{1,j}\downarrow_{2,j}\uparrow_{3,j}\downarrow_{4,j}\uparrow_{5,j}\uparrow_{6,j}\rangle \nonumber \\
-|\!\!\uparrow_{1,j}\uparrow_{2,j}\downarrow_{3,j}\downarrow_{4,j}\downarrow_{5,j}\uparrow_{6,j}\rangle 
\!\!\!&-&\!\!\!|\!\!\uparrow_{1,j}\downarrow_{2,j}\downarrow_{3,j}\uparrow_{4,j}\downarrow_{5,j}\uparrow_{6,j}\rangle \nonumber \\
-|\!\!\downarrow_{1,j}\uparrow_{2,j}\uparrow_{3,j}\uparrow_{4,j}\downarrow_{5,j}\downarrow_{6,j}\rangle 
\!\!\!&-&\!\!\!|\!\!\downarrow_{1,j}\uparrow_{2,j}\downarrow_{3,j}\uparrow_{4,j}\uparrow_{5,j}\downarrow_{6,j}\rangle \Bigr), \nonumber \\
\label{SH}
\end{eqnarray}
while the latter state vector refers to the spin-$\frac{1}{2}$ Heisenberg square in the singlet-tetramer state explicitly given by Eq. (\ref{ST}). The singlet tetramer-hexamer state has a spontaneously broken symmetry on behalf of a regular alternation of singlets, which are being alternatively formed on octahedrons (hexamers) and square plaquettes (tetramers) as schematically illustrated in Fig. \ref{fig4}.

\subsection{Exact diagonalization} 

To avoid a danger of overlooking some higher-period ground state(s) of the spin-$\frac{1}{2}$ Heisenberg octahedral chain we have employed the numerical ED method based on the Lanczos algorithm for the 
finite-size chains with $L=20,25,30,35,40$ spins ($N=4,5,6,7,8$ unit cells) imposing periodic
boundary conditions. The ED data can be regarded as a useful benchmark for the numerical data obtained from DMRG simulations of the effective ferrimagnetic mixed-spin Heisenberg chains, because any substantial discrepancy between these numerical results 
would indicate that some higher-period quantum ground state was disregarded. The ED data for various chain lengths show very small finite-size
effects and fit well to the DMRG results, see below.

Beside this, we have also performed the full ED of the finite-size spin-$\frac{1}{2}$ Heisenberg octahedral chain with up to $L=20$ spins ($N=4$ unit cells) in order to verify reliability of the developed localized-magnon approach for a description of the low-temperature thermodynamics in the highly frustrated parameter space $J_2 > 2 J_1$. To this end, we have adapted for the full ED calculations the subroutines from the Spinpack 
project.\cite{rich10,schu10} 

\section{Results and discussion}
\label{sec:result}

In this section, we will perform a comprehensive analysis of the most interesting results for the ground state, magnetization process and low-temperature thermodynamics of the spin-$\frac{1}{2}$ Heisenberg octahedral chain.    

\subsection{Ground-state phase diagrams}

\begin{figure}
\begin{center}
\hspace*{-1.8cm}
\includegraphics[width=0.68\textwidth]{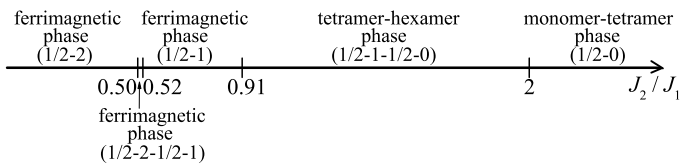}
\end{center}
\vspace{-0.7cm}
\caption{The zero-field ground-state phase diagram of the spin-$\frac{1}{2}$ Heisenberg octahedral chain. 
The numbers in square brackets determine  within a given ground state the total spin on monomeric sites and square plaquettes.
Note that the ferrimagnetic phase (1/2-2-1/2-1) and the monomer-tetramer phase (1/2-1-1/2-0) break the translational symmetry.}
\label{fig5}
\end{figure}

Let us begin with the analysis of ground state at zero magnetic field. The zero-field ground-state phase diagram is schematically depicted in Fig. \ref{fig5} and it totally involves five different ground states (three quantum ferrimagnetic states, a singlet tetramer-hexamer state and a monomer-tetramer state) depending on a relative strength of two considered coupling constants. At small values of $J_2/J_1<0.50$ the ground state of the spin-$\frac{1}{2}$ Heisenberg octahedral chain can be described as the quantum ferrimagnetic state of the effective mixed spin-($\frac{1}{2},2$) Heisenberg chain \cite{ivanov98} with the energy eigenvalue (\ref{ee122}). Another quantum ferrimagnetic ground state with a doubled period of the magnetic unit cell relates to the lowest-energy eigenstate (\ref{eeeffd}) of the effective mixed spin-($\frac{1}{2},2,\frac{1}{2},1$) Heisenberg chain, which emerges just in a relatively narrow parameter region $J_2/J_1 \in (0.50, 0.52)$. The last quantum ferrimagnetic ground state of the spin-$\frac{1}{2}$ Heisenberg octahedral chain stems from the lowest-energy eigenstate (\ref{ee121}) of the effective mixed spin-($\frac{1}{2},1$) Heisenberg chain, which has the lowest energy in the parameter region $J_2/J_1 \in (0.52, 0.91)$. The bipartite nature of the effective mixed spin-($\frac{1}{2},2$) and spin-($\frac{1}{2},1$) Heisenberg chains implies that two related quantum ground states can be identified with the conventional ferrimagnetic phases of Lieb-Mattis type.\cite{lieb62} The same conclusion could be also inferred for the third quantum ferrimagnetic ground state even though the four-sublattice character of the effective mixed spin-($\frac{1}{2},2,\frac{1}{2},1$) Heisenberg chain precludes the simple argumentation on the grounds of Lieb-Mattis theorem.\cite{lieb62}  

Apart from the three aforementioned quantum ferrimagnetic ground states one also encounters two quantum ground states underlying fragmentation at square plaquettes in the singlet-tetramer state (\ref{ST}). The singlet tetramer-hexamer ground state (\ref{SHT}) with a spontaneously broken symmetry (Fig. \ref{fig4}) is the lowest-energy eigenstate of the spin-$\frac{1}{2}$ Heisenberg octahedral chain at moderate values of the coupling constants $J_2/J_1 \in (0.91, 2)$. Finally, the ground state of the spin-$\frac{1}{2}$ Heisenberg octahedral chain is the fully fragmentized monomer-tetramer state (\ref{MT}) whenever the intra-plaquette coupling is at least twice as strong as the monomer-plaquette coupling $J_2/J_1 > 2$.

\begin{figure}
\begin{center}
\includegraphics[width=0.45\textwidth]{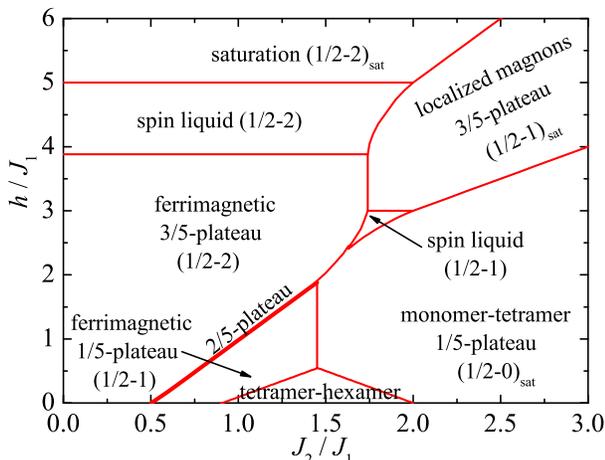}
\end{center}
\vspace{-0.8cm}
\caption{(Color online) The ground-state phase diagram of the spin-$\frac{1}{2}$ Heisenberg octahedral chain in the $J_2/J_1 - h/J_1$ plane. A thick line schematically shows the quantum ferrimagnetic ground state, which originates from the effective mixed spin-($\frac{1}{2},2,\frac{1}{2},1$) Heisenberg chain and is stable in a very narrow interval of the magnetic fields.}
\label{fig6}
\end{figure}

It is quite clear that the three quantum ferrimagnetic ground states related to the lowest-energy eigenstates of the effective mixed spin-($\frac{1}{2},2$), spin-($\frac{1}{2},2,\frac{1}{2},1$) and spin-($\frac{1}{2},1$) Heisenberg chains should be manifested in zero-temperature magnetization curves as intermediate plateaus at $3/5$, $2/5$ and $1/5$ of the saturation magnetization, respectively. In addition, the singlet tetramer-hexamer ground state (\ref{SHT}) should be responsible for a zero magnetization plateau, while the monomer-tetramer state (\ref{MT}) affords another $1/5$-plateau state due to a full polarization of the monomeric spins. The overall ground-state phase diagram of the spin-$\frac{1}{2}$ Heisenberg octahedral chain elucidating the effect of external magnetic field is displayed in Fig. \ref{fig6} in the $J_2/J_1 - h/J_1$ plane. It is evident that the highly-frustrated parameter region $J_2/J_1 > 2$ of the ground-state phase diagram is fully consistent with the rigorous theoretical predictions of the monomer-tetramer state (\ref{MT}) and the localized-magnon state (\ref{LM}) gained in Sects. \ref{vm} and \ref{lmgs} by making use of the variational procedure and localized-magnon approach. Moreover, one finds that the intermediate 3/5-plateau due to the localized-magnon state (\ref{LM}) with a single magnon trapped on each square plaquette can alternatively be interpreted as the saturated state of the effective ferrimagnetic mixed spin-($\frac{1}{2},1$) Heisenberg chain. 

On the other hand, the ground-state phase diagram of the spin-$\frac{1}{2}$ Heisenberg octahedral chain is much more intricate in the parameter region $J_2/J_1 < 2$ because of presence of two different spin-liquid ground states with short-range correlations but without any spontaneously broken symmetry.\cite{lhui02,bale10,misg11} If the relative strength of the coupling constants is sufficiently weak, i.e. $J_2/J_1<0.50$, the ground state of the spin-$\frac{1}{2}$ Heisenberg octahedral chain entirely follows from the lowest-energy eigenstates (\ref{ee122}) of the effective mixed spin-($\frac{1}{2},2$) Heisenberg chain. Consequently, the intermediate 3/5-plateau due to the Lieb-Mattis ferrimagnetic ($\frac{1}{2}-2$) ground state breaks down at a field-driven quantum critical point towards the gapless spin-liquid ($\frac{1}{2}-2$) ground state. The similar scenario can be detected in the parameter space $J_2/J_1 \in (0.50, 0.52)$ except that the tiny 2/5-plateau emerges at low enough magnetic fields due to the quantum ferrimagnetic ($\frac{1}{2}-2-\frac{1}{2}-1$) ground state stemming from the lowest-energy eigenstate (\ref{eeeffd}) of the effective mixed spin-($\frac{1}{2},2,\frac{1}{2},1$) Heisenberg chain. In the parameter region $J_2/J_1 \in (0.52, 0.91)$ the Lieb-Mattis ferrimagnetic ($\frac{1}{2}-1$) ground state arising out from the effective mixed spin-($\frac{1}{2},1$) Heisenberg chain is responsible for the extra 1/5-plateau at sufficiently low magnetic fields. Last but not least, the zero magnetization plateau reflecting the singlet tetramer-hexamer ground state (\ref{SHT}) emerges at moderate values of the coupling constants $J_2/J_1 \in (0.91, 2)$. Within this parameter region, the 1/5-plateau emerging above the zero magnetization plateau either corresponds to the Lieb-Mattis quantum ferrimagnet ($\frac{1}{2}-1$) or the monomer-tetramer phase (\ref{MT}) depending on whether a relative size of the coupling constants is smaller or greater than the threshold value $J_2/J_1 = 1.45$. Analogously, the intermediate 3/5-plateau is either due to the Lieb-Mattis quantum ferrimagnet ($\frac{1}{2}-2$) or the localized-magnon ground state (\ref{LM}) depending on whether a relative strength of the coupling constants is smaller or greater than the threshold value $J_2/J_1 = 1.74$. Finally, another gapless spin-liquid ($\frac{1}{2}-1$) ground state can be found in the parameter region $J_2/J_1 \in (1.63, 2)$ and $h/J_1 \lesssim 3$ on behalf of the lowest-energy eigenstate of the effective mixed spin-($\frac{1}{2},1$) Heisenberg chain. It is worthwhile to remark that all displayed phase boundaries represent discontinuous quantum phase transitions with exception of three horizontal borders related to continuous quantum phase transitions.

\subsection{Zero-temperature magnetization curves}

\begin{figure*}
\begin{center}
\includegraphics[width=0.45\textwidth]{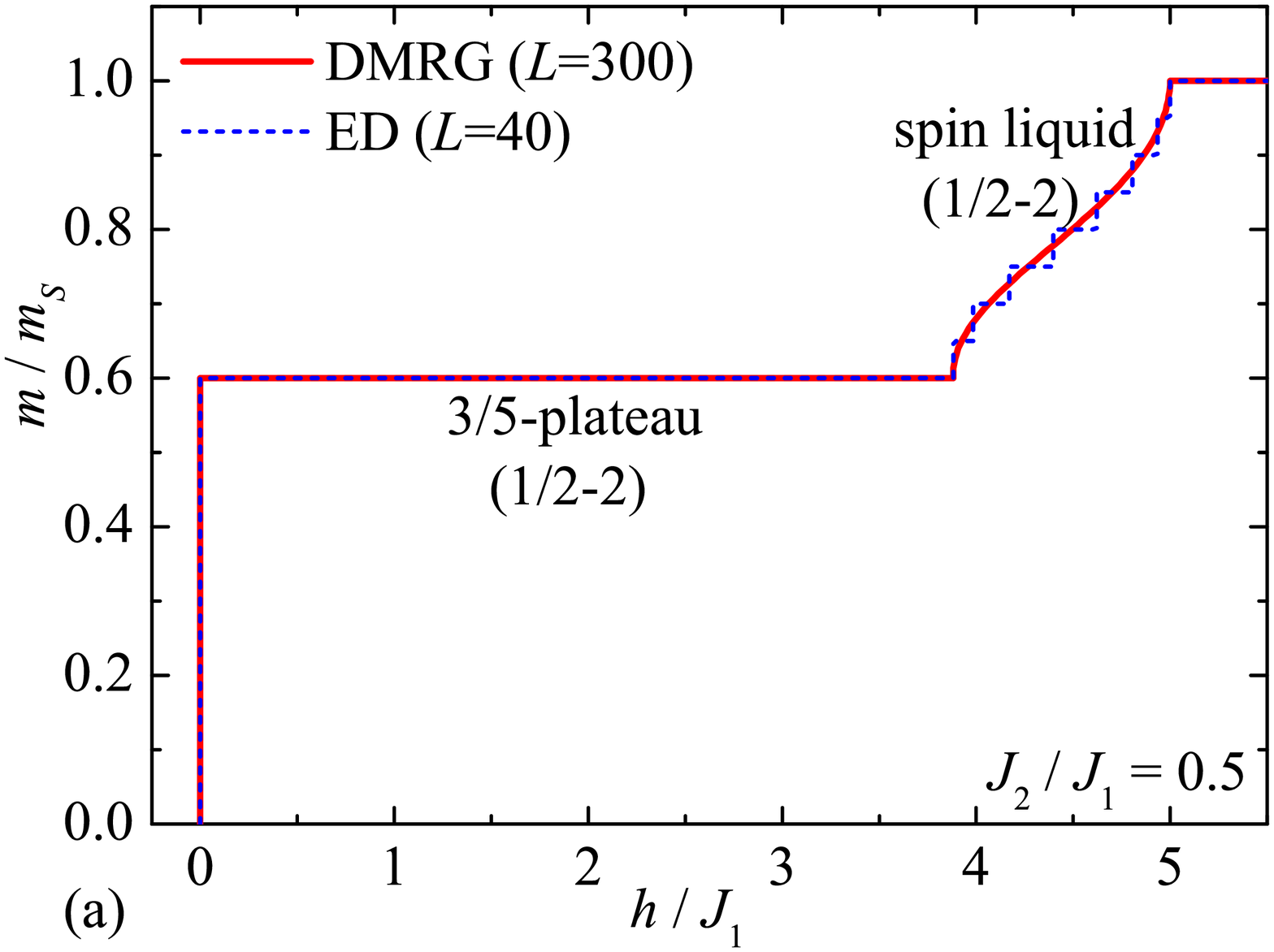}
\hspace{0.5cm}
\includegraphics[width=0.45\textwidth]{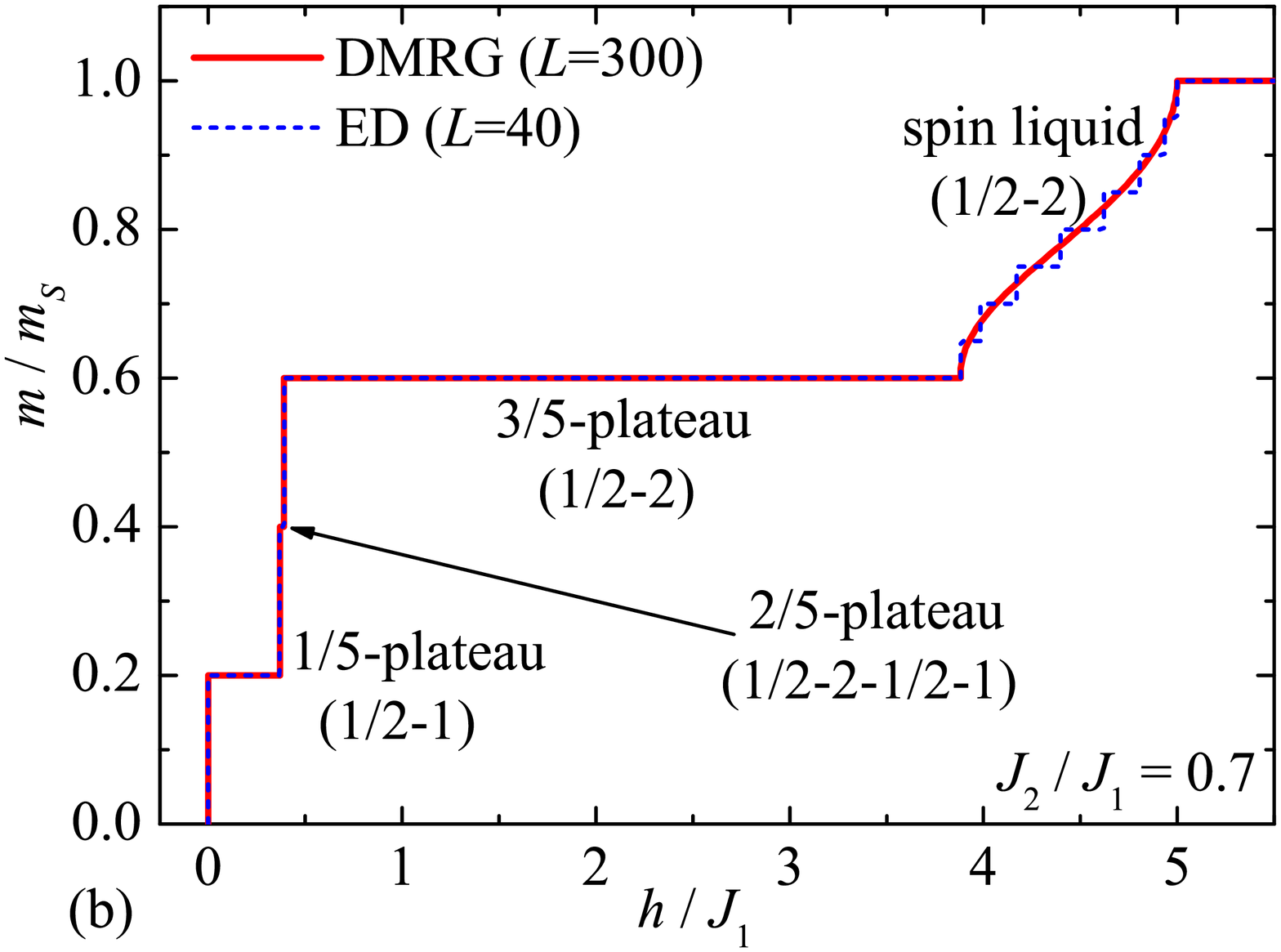}
\includegraphics[width=0.45\textwidth]{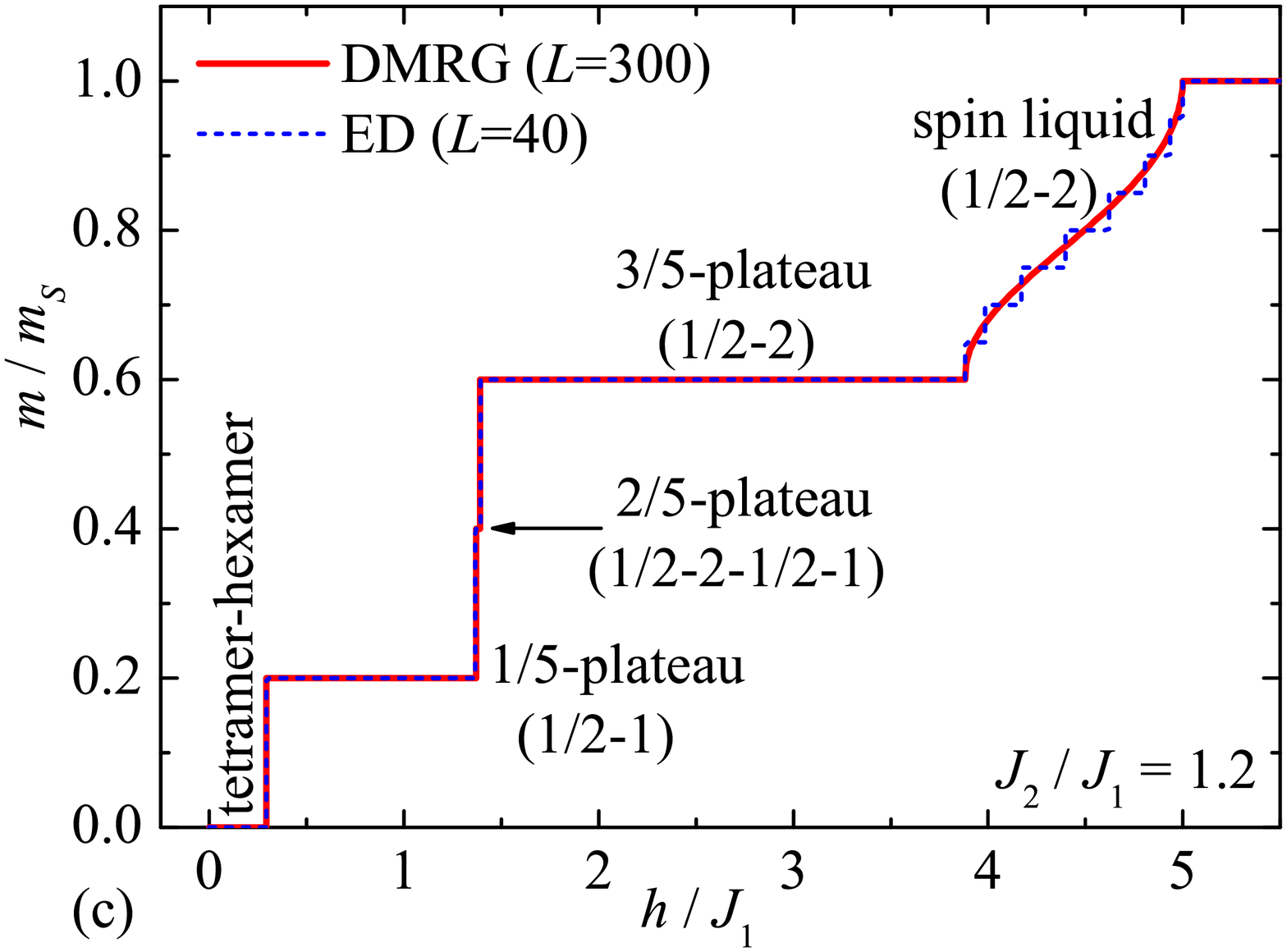}
\hspace{0.5cm}
\includegraphics[width=0.45\textwidth]{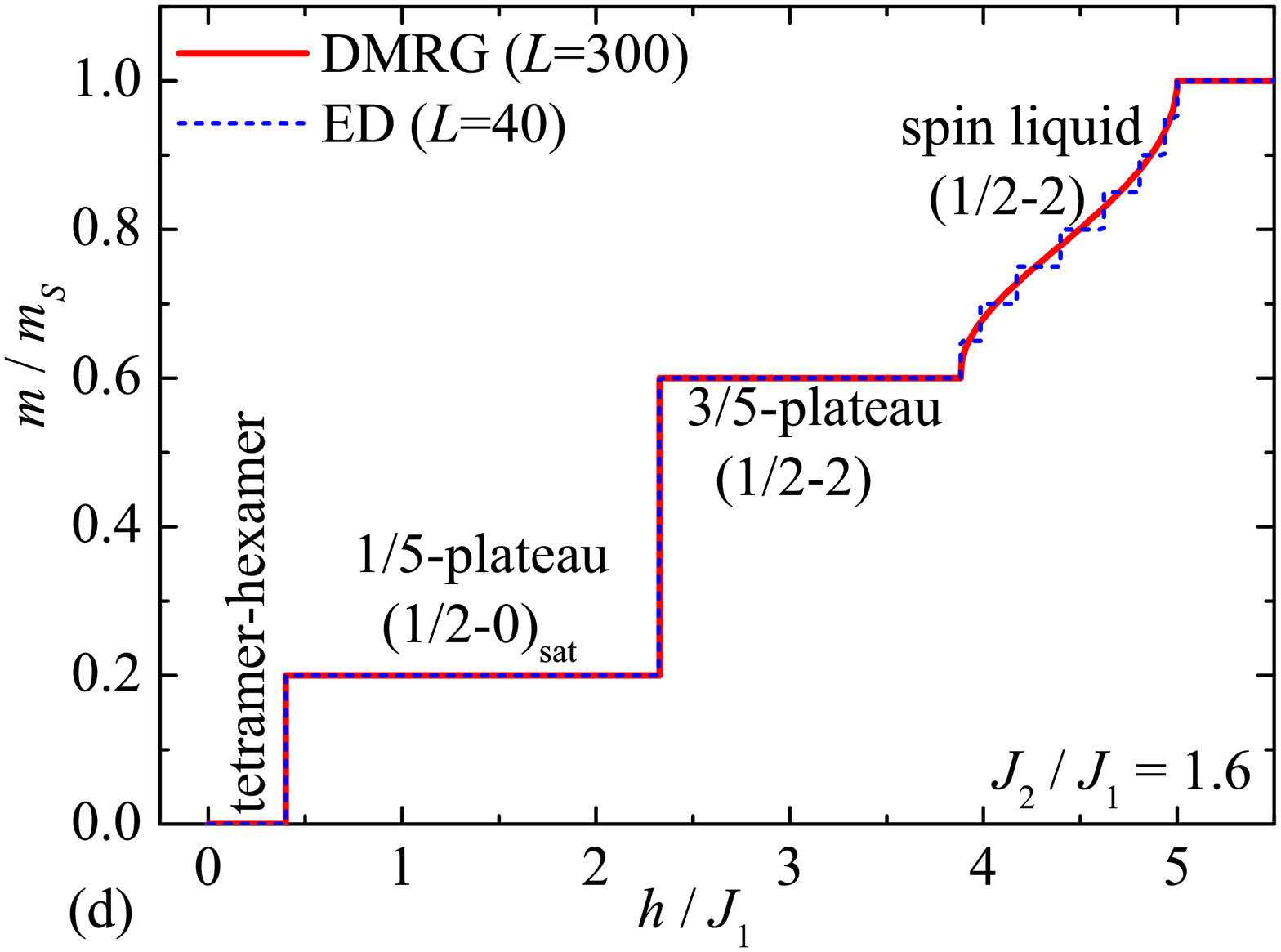}
\includegraphics[width=0.45\textwidth]{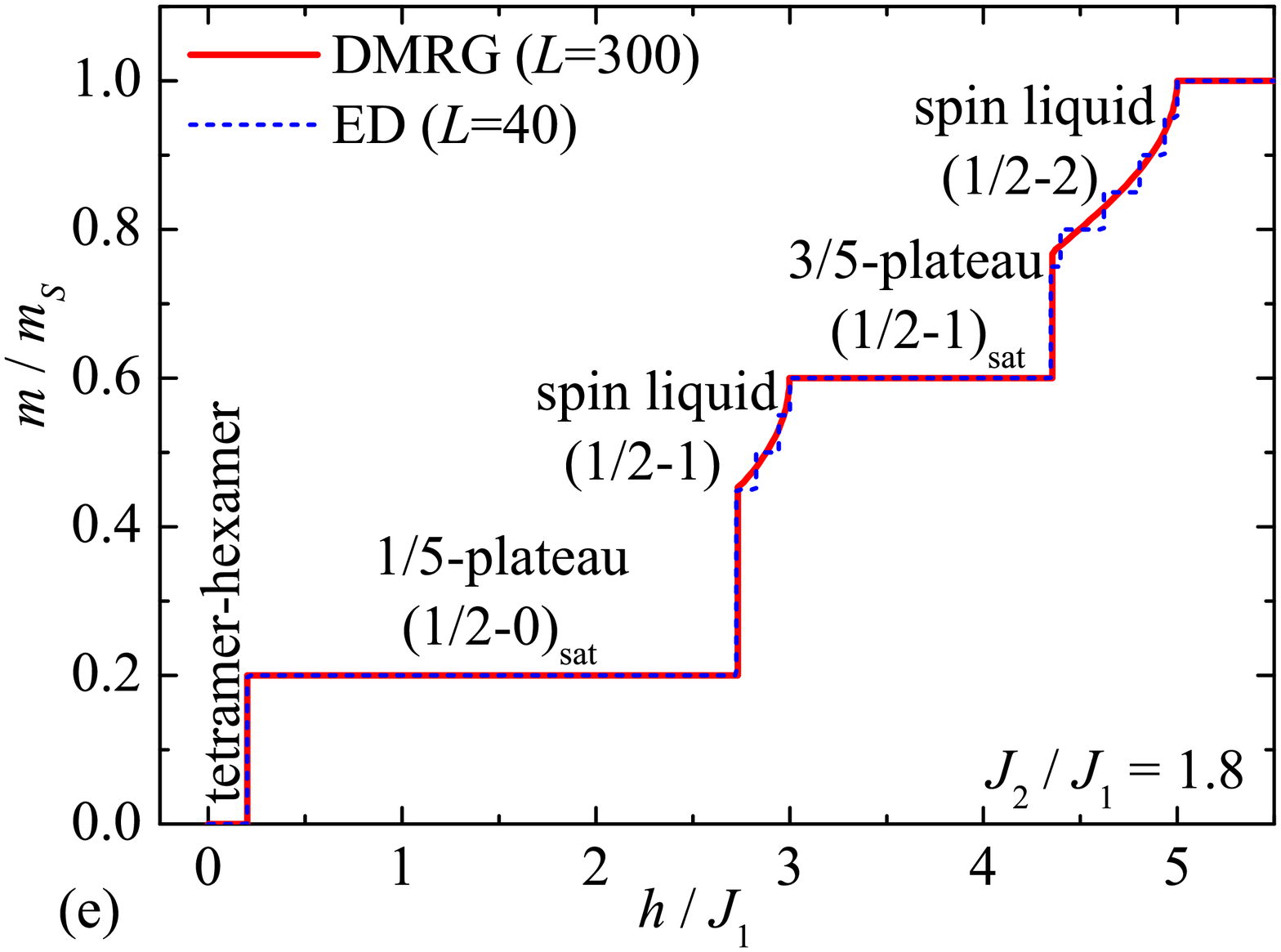}
\hspace{0.5cm}
\includegraphics[width=0.45\textwidth]{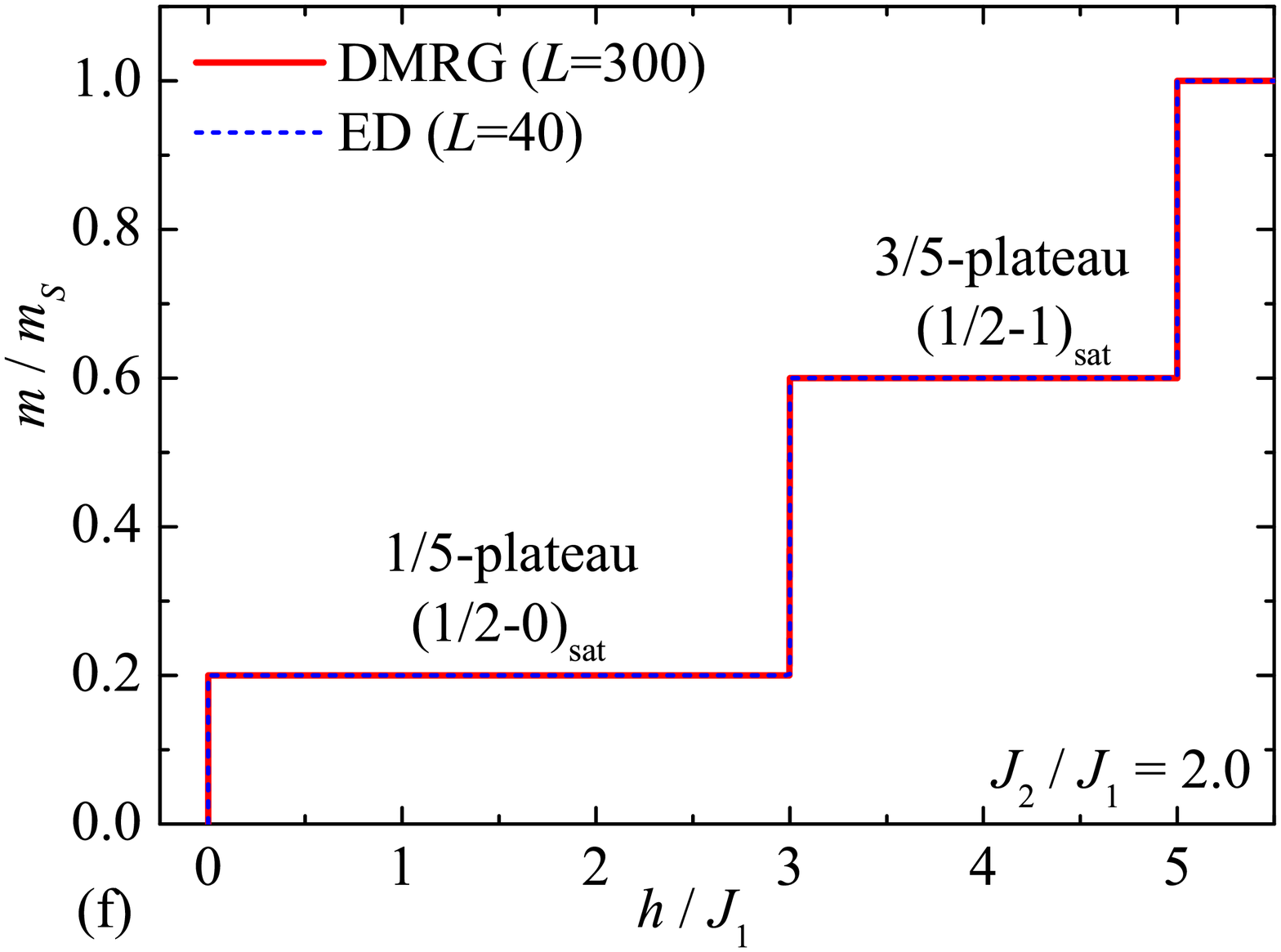}
\end{center}
\vspace{-0.8cm}
\caption{(Color online) A few typical zero-temperature magnetization curves of the spin-$\frac{1}{2}$ Heisenberg octahedral chain. Thick solid lines show numerical results based on DMRG calculations for the finite-size chain of $L=300$ spins ($N=60$ unit cells) and thin broken lines illustrate numerical ED results for the  finite-size chain of $L=40$ spins ($N=8$ unit cells) by assuming six different values of the interaction ratio: (a) $J_2/J_1 = 0.5$; (b) $J_2/J_1 = 0.7$; (c) $J_2/J_1 = 1.2$; (d) $J_2/J_1 = 1.6$; (e) $J_2/J_1 = 1.8$; (f) $J_2/J_1 = 2.0$.}
\label{fig7}
\end{figure*}

To illustrate the great diversity of possible scenarios of the magnetization process of the spin-$\frac{1}{2}$ Heisenberg octahedral chain we have depicted in Fig. \ref{fig7} a few typical examples of zero-temperature magnetization curves. The magnetization data obtained from the DMRG simulations of the effective mixed-spin Heisenberg chains are shown in Fig. \ref{fig7} by thick solid lines and for a comparison, we have also plotted the magnetization data from ED method by thin broken lines. As one can see, there is in general perfect coincidence between the zero-temperature magnetization curves from DMRG and ED calculations as far as the height and width of all intermediate magnetization plateaus is concerned. The only noticeable difference is thus attributable to a finite-size effect of the ED method, which is most clearly seen within both gapless spin-liquid ground states manifested through a continuous rise of the magnetization with the magnetic field. 

It should be also stressed that the displayed magnetization curves corroborate correctness of the established ground-state phase diagram. The magnetization curve shown in Fig. \ref{fig7}(a) demonstrates two continuous quantum phase transitions, which occur at a rise and fall of the intermediate 3/5-plateau. The magnetization curve plotted in Fig. \ref{fig7}(b) verifies uprise of 1/5- and 2/5-plateau, which is achieved upon small strengthening of the relative ratio $J_2/J_1$. The emergence of zero magnetization plateau on account of the singlet tetramer-hexamer ground state (\ref{SHT}) can be seen in the magnetization curve shown in Fig. \ref{fig7}(c). Furthermore, Fig. \ref{fig7}(d) illustrates the magnetization curve with a change in character of 1/5-plateau as well as disappearance of the subtle 2/5-plateau. The magnetization curve plotted in Fig. \ref{fig7}(e) demonstrates change over the character of 3/5-plateau as well as two emergent magnetization jumps towards the spin-liquid ground states. Finally, Fig. \ref{fig7}(f) displays typical magnetization curve in the highly frustrated region $J_2/J_1>2$ involving three abrupt magnetization jumps.

\subsection{Low-temperature thermodynamics}

\begin{figure*}
\begin{center}
\includegraphics[width=0.45\textwidth]{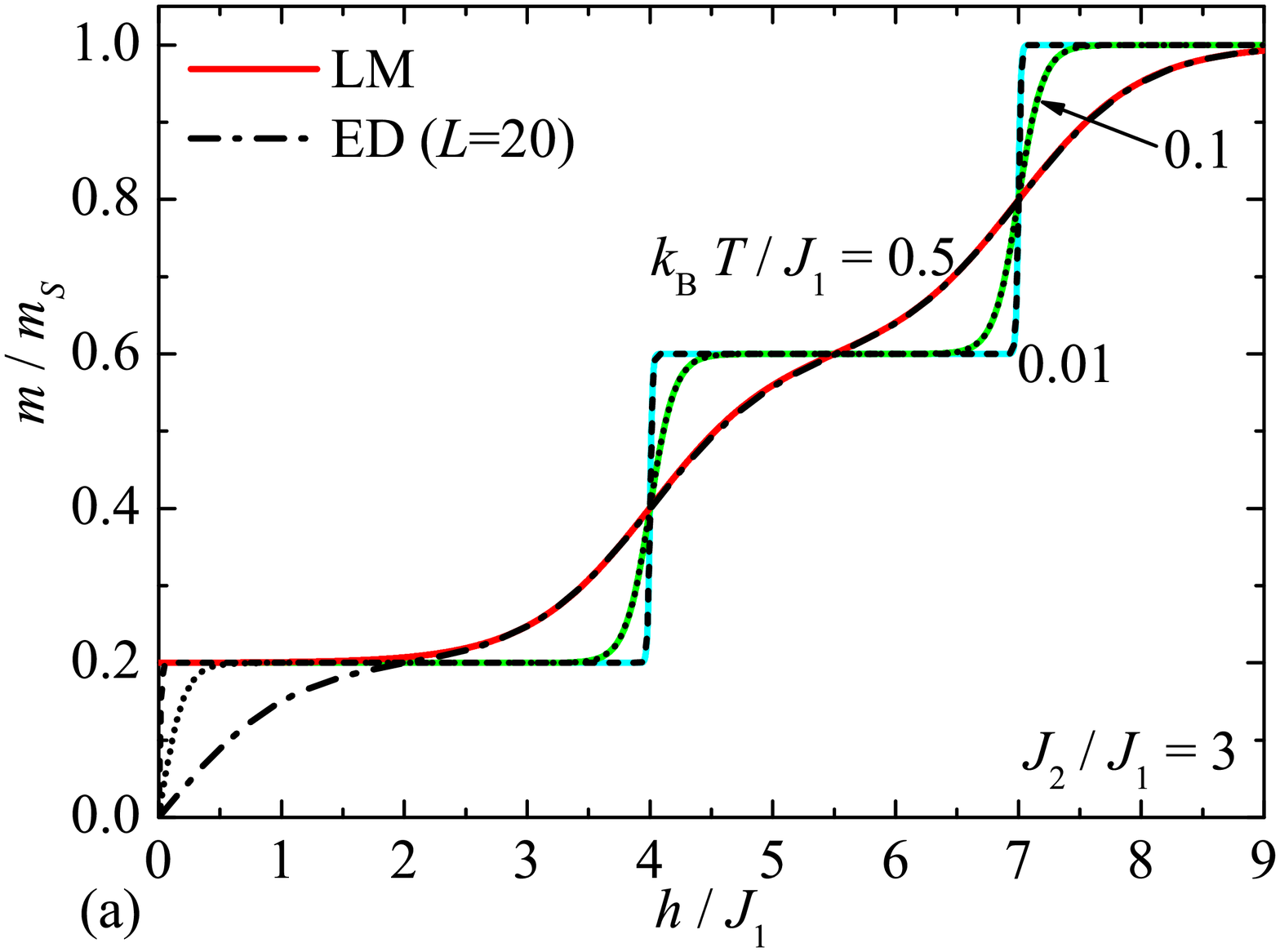}
\hspace{0.5cm}
\includegraphics[width=0.45\textwidth]{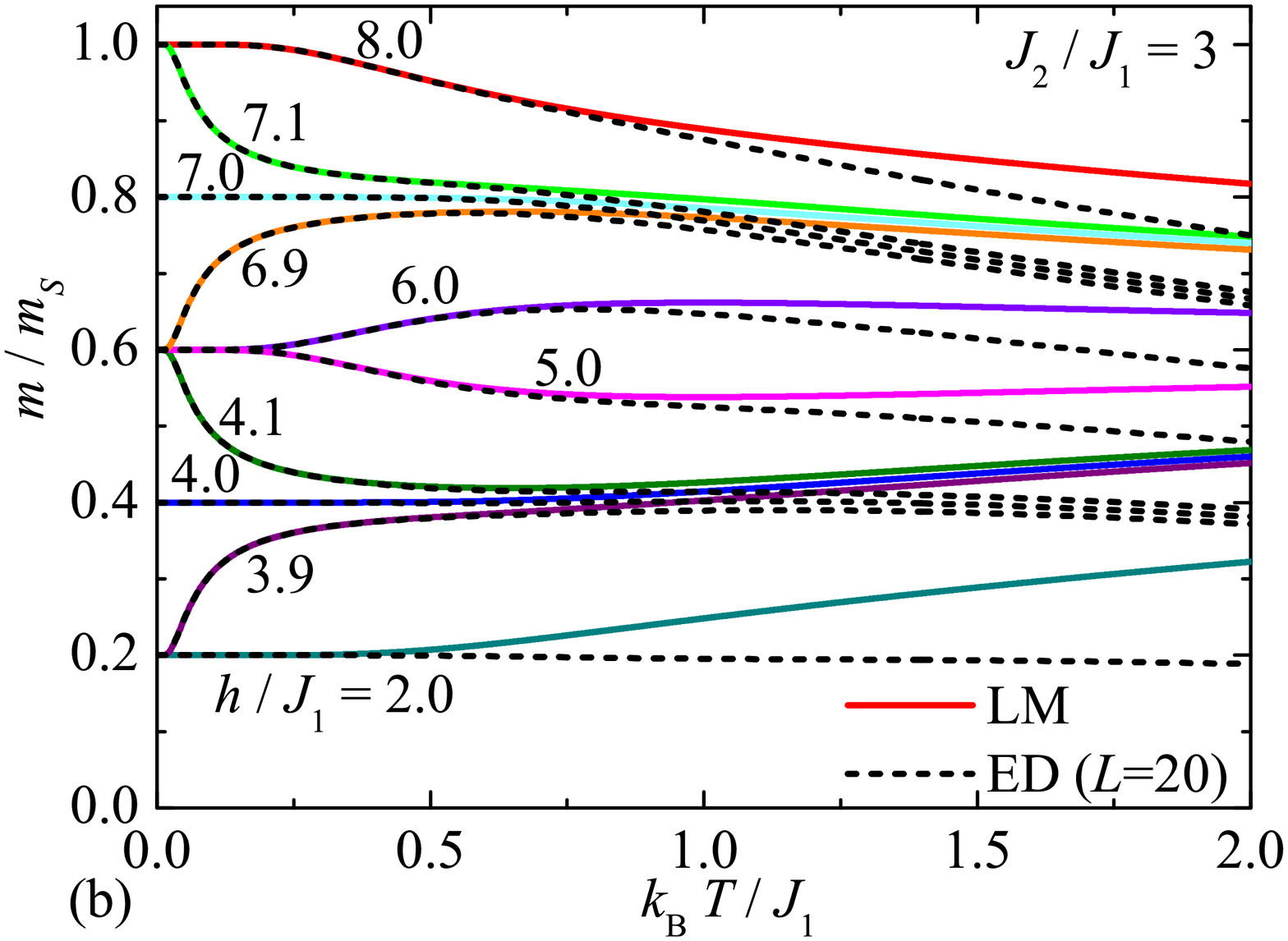}
\includegraphics[width=0.45\textwidth]{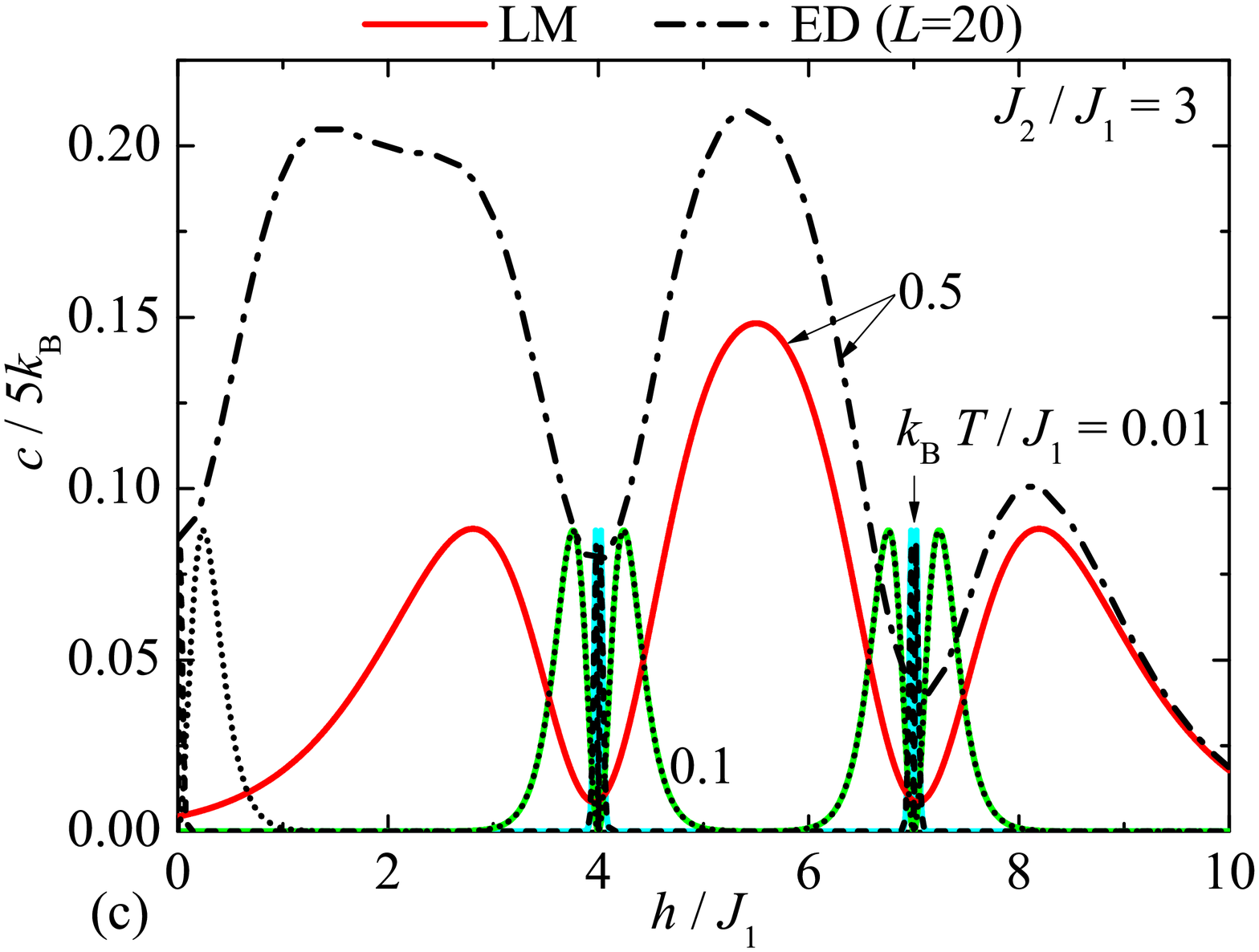}
\hspace{0.5cm}
\includegraphics[width=0.45\textwidth]{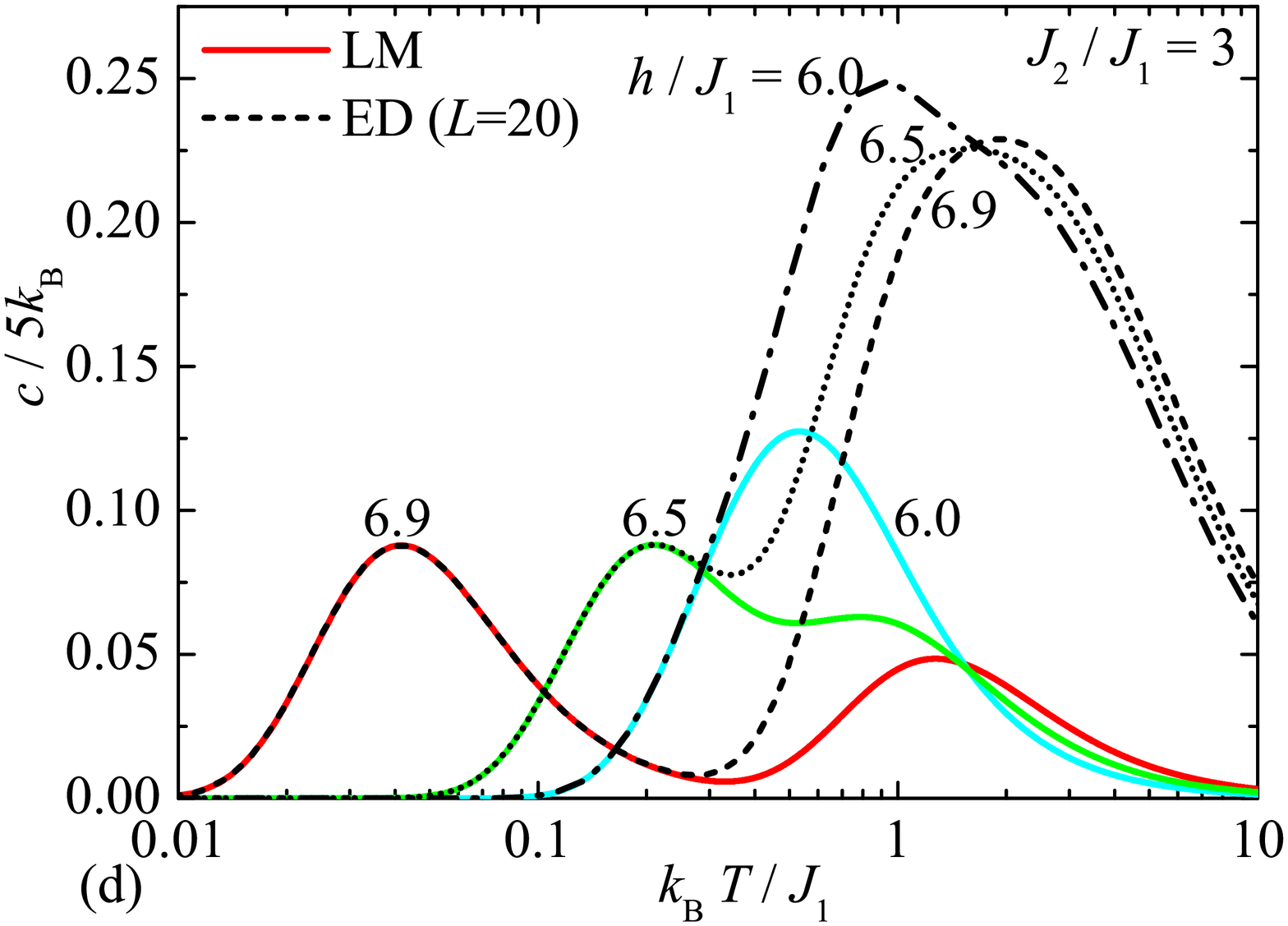}
\end{center}
\vspace{-0.8cm}
\caption{(Color online) The magnetization (upper panel) and specific heat (lower panel) of the spin-$\frac{1}{2}$ Heisenberg octahedral chain as a function of the magnetic field and temperature for the fixed value of the interaction ratio $J_2/J_1 = 3$. Solid lines follow from Eqs. (\ref{lmmag}) and (\ref{lmsh}) derived by means of the localized-magnon approach, while broken lines of different styles illustrate the full ED data for the finite-size chain of $L = 20$ spins ($N=4$ unit cells).}
\label{fig8}
\end{figure*}

Next, let us proceed to a discussion of the most interesting results for the 
low-temperature thermodynamics of the spin-$\frac{1}{2}$ Heisenberg octahedral chain, which were obtained within the framework of the localized-magnon approach elaborated in Sect. \ref{lmt}. 
It is worthwhile to recall, however, that the localized-magnon theory provides reasonable results only in the highly-frustrated parameter region $J_2/J_1 > 2$, where the many-magnon states constructed from the lowest-energy one-magnon (\ref{om1}) and two-magnon (\ref{ST}) localized states are the 
most relevant ones for a proper description of the low-temperature thermodynamics.  

For the sake of comparison and verification, the magnetization and specific heat of the spin-$\frac{1}{2}$ Heisenberg octahedral chain obtained from the localized-magnon approach are plotted in 
Fig. \ref{fig8} along with the analogous data for the finite-size chain of $L = 20$ spins obtained within the full ED method. 
Let us first emphasize two interesting features, namely (i) the
extra low-temperature peak in the specific heat for magnetic fields slightly
below the saturation field, cf. Fig.~\ref{fig8}(d), and, (ii)
the noticeable increase of the magnetization with growing temperature 
 for $h/J_1=3.9$ and $6.9$, cf. Fig.~\ref{fig8}(b).
Both unconventional features are related to the flat bands in the
one-magnon excitations, cf. Fig.~\ref{fig3}.
 
It can be understood from Fig.~\ref{fig8} that the localized-magnon theory provides a plausible description of the low-temperature magnetization and thermodynamics whenever the external magnetic field drives the investigated spin system above a midpoint of the lowest intermediate 1/5-plateau appearing due to the monomer-tetramer ground state (\ref{MT}). Under this circumstance, the magnetization calculated according to Eq. (\ref{lmmag}) exhibits for $J_2/J_1 = 3$ a perfect agreement with the ED data on assumption that the temperature is below $k_{\rm B} T/J_1 \lesssim 0.5$. Compared to this, the derived formula (\ref{lmsh}) for the specific heat affords a reliable description of the numerical ED data for the same value of the interaction ratio $J_2/J_1 = 3$ just for much smaller temperatures $k_{\rm B} T/J_1 \lesssim 0.2$. It should be pointed out, however, that the height and position of 
low-temperature maximum of the heat capacity is in an excellent accordance with the numerical ED data provided that the magnetic field is fixed sufficiently close to the saturation field or the field-driven transition 
between $1/5$- and $3/5$-plateaus.

\section{Conclusion}
\label{sec:conc}

The present work deals with the ground-state phase diagram, the magnetization process and the low-temperature thermodynamics of the spin-$\frac{1}{2}$ Heisenberg octahedral chain, which has been treated by means of various analytical and numerical techniques. It has been demonstrated that the highly-frustrated parameter region of the ground-state phase diagram can be rigorously found on the grounds of the variational principle and the localized-magnon approach, which provide an exact evidence for the monomer-tetramer (\ref{MT}) and localized-magnon (\ref{LM}) ground states at low and high magnetic fields, respectively. On the other hand, the remaining part of the ground-state phase diagram was established through the numerical data gained from DMRG simulations of the effective mixed-spin Heisenberg chains, which were additionally corroborated by ED data exploiting Lanczos algorithm. It has been verified that the spin-$\frac{1}{2}$ Heisenberg octahedral chain exhibits in this parameter space an unexpected diversity of intriguing ground states including three different quantum ferrimagnetic phases, two spin-liquid phases as well as the unconventional singlet tetramer-hexamer phase (\ref{SHT}). 

The notable diversity of available quantum ground states can also be regarded as a primary reason for astonishing versatility of zero-temperature magnetization curves. As a matter of fact, the magnetization curves of the spin-$\frac{1}{2}$ Heisenberg octahedral chain may involve intermediate plateaus at zero, one-fifth, two-fifth and three-fifth of the saturation magnetization in addition to two gapless spin-liquid regimes with continuously varying magnetization. The field-driven quantum phase transition between individual ground states can therefore have either character of a first-order phase transition accompanied with a discontinuous magnetization jump or of a second-order phase transition accompanied with a continuous rise of the magnetization. 

Last but not least, we have developed the modified localized-magnon theory accounting for the lowest-energy one-magnon and two-magnon states, which are essential for a design of the many-magnon states relevant for a proper description of low-temperature thermodynamics of the spin-$\frac{1}{2}$ Heisenberg octahedral chain. The validity and exactness of the developed localized-magnon approach in the highly-frustrated parameter region has been verified through a direct comparison with the full ED data. It has been found that the localized-magnon theory gives plausible estimate of thermodynamic quantities at low up to moderate temperatures whenever the magnetic field drives the investigated quantum spin system above a midpoint of the intermediate 1/5-plateau due to the monomer-tetramer (\ref{MT}) phase. Our future goal is to extend a validity of the localized-magnon approach down to zero field, because a steep variation of the magnetization at low magnetic fields is of great potential applicability for low-temperature magnetic refrigeration achieved through the enhanced magnetocaloric 
effect.\cite{zhit03,zhit04,schnack07,schnack13}           

\begin{acknowledgments}
This work was financially supported by the grant of The Ministry of Education, Science, Research and Sport of the Slovak Republic under the contract No. VEGA 1/0043/16 and by the grant of the Slovak Research and Development Agency under the contract No. APVV-14-0073. O.D. and J.R. acknowledge the support by the Deutsche Forschungsgemeinschaft (project RI615/21-2). O.D. was partially supported by Project FF-30F (No. 0116U001539) from the Ministry of Education and Science of Ukraine.
\end{acknowledgments}

\end{document}